\documentclass[%
 reprint,
 superscriptaddress,
nofootinbib,
 amsmath,amssymb,smsymb
 aps, apl, prd,
 nofo,
 otinbib,
 floatfix
]{revtex4-2}

\usepackage[mathlines]{lineno}      

\usepackage{graphicx}                           
\usepackage[justification=Justified]{caption}   
\usepackage{subcaption}					    

\usepackage{rotating}                       
\usepackage{booktabs}
\usepackage{tabularx}
\usepackage{ragged2e}

\usepackage{dcolumn}                        
\usepackage{xcolor}
\usepackage{hyperref}                       

\usepackage{amsmath}						
\usepackage{bm}                             
\usepackage{amssymb}						
\usepackage{mathtools}						

\usepackage{siunitx}						
\usepackage{mathrsfs}						
\usepackage{latexsym}						
\usepackage{revsymb}						

\usepackage{scalerel}                       

\usepackage{soul}
\usepackage{natbib}
\usepackage{url}
\usepackage{textcase}
\usepackage{orcidlink}

\begin{document}

\preprint{APS/123-QED}

\title{Constraints on the amplitude of gravitational wave echoes from black hole ringdown using minimal assumptions}

\author{Andrea Miani \orcidlink{0000-0001-7737-3129}}
    \email[Correspondence email adress: ]{andrea.miani@unitn.it}
    \affiliation{Università di Trento, Dipartimento di Fisica, I-38123 Povo, Trento, Italy}
    \affiliation{INFN, Trento Institute for Fundamental Physics and Applications, I-38123 Povo, Trento, Italy}
\author{Claudia Lazzaro \orcidlink{0000-0001-5993-3372}}
    \affiliation{Università di Padova, Dipartimento di Fisica, I-35121 Padova, Italy}
    \affiliation{INFN, Sezione di Padova, I-35131 Padova, Italy}
\author{Giovanni Andrea Prodi \orcidlink{0000-0001-5256-915X}}
    \affiliation{Università di Trento, Dipartimento di Matematica, I-38123 Povo, Trento, Italy}
    \affiliation{INFN, Trento Institute for Fundamental Physics and Applications, I-38123 Povo, Trento, Italy}

\author{Shubhanshu Tiwari \orcidlink{0000-0003-1611-6625}}
    \affiliation{ Physik-Institut, University of Zurich, Winterthurerstrasse 190, 8057 Zurich, Switzerland}
\author{Marco Drago \orcidlink{0000-0002-3738-2431}}
    \affiliation{Università di Roma, Dipartimento di Fisica, I-00133 Roma, Italy}
    \affiliation{INFN, Sezione di Roma, I-00133 Roma, Italy}
\author{Edoardo Milotti \orcidlink{0000-0001-7348-9765}}
    \affiliation{Università di Trieste, Dipartimento di Fisica, I-34127 Trieste, Italy}
    \affiliation{INFN, Sezione di Trieste, I-34127 Trieste, Italy}

\author{Gabriele Vedovato \orcidlink{0000-0001-7226-1320}}
    \affiliation{Università di Padova, Dipartimento di Fisica, I-35121 Padova, Italy}
    \affiliation{INFN, Sezione di Padova, I-35131 Padova, Italy}


\date{\today}

\begin{abstract}
Gravitational wave echoes may appear following a compact binary coalescence if the remnant is an ``exotic compact object'' (ECO).
ECOs are proposed alternatives to the black holes of Einstein's general relativity theory and are predicted to possess reflective boundaries.
This work reports a search for gravitational wave transients (GWTs) of generic morphology occurring shortly after ($\lesssim \SI{1}{\second}$) binary black hole (BBH) mergers, therefore targeting all gravitational wave echo models.
We investigated the times after the ringdown for the higher signal-to-noise ratio BBHs within the public catalog GWTC-3 by the LIGO-Virgo-KAGRA collaborations (LVK).
Our search is based on the coherent WaveBurst pipeline, widely used in generic searches for GWTs by the LVK, and deploys new methods to enhance its detection performances at low signal-to-noise ratios.
We employ Monte Carlo simulations for estimating the detection efficiency of the search and determining the statistical significance of candidates.
We find no evidence of previously undetected GWTs and our loudest candidates are morphologically consistent with known instrumental noise disturbances.
Finally, we set upper limits on the amplitude of GW echoes for single BBH mergers.
\end{abstract}

\maketitle


\section{Introduction}
\label{sec:introduction}
The LIGO \cite{adLIGO} and Virgo \cite{adVIRGO} observatories have successfully detected about 90 gravitational wave transients (GWTs) in past observing runs \cite{gwtc1,gwtc2,gwtc3}, all associated to compact binary coalescences (CBCs).
More than $90\%$ of these GWTs are identified as generated by the merger of binary black hole (BBH) systems.
Recently, this worldwide observatory has expanded to include the KAGRA detector \cite{kagra}.
A new observing run is currently ongoing, and low latency alerts of more CBC GWTs are being publicly released \cite{gracedb}.
Investigating the black hole (BH) nature through GW astronomy is therefore a very hot topic in fundamental physics, especially in view of the so-called BH information paradox \cite{information_paradox}.
The LIGO-Virgo-KAGRA collaboration (LVK) already published several results of tests of the general relativity theory (GR) \cite{GR,tgrGW150914,tgr1,tgr2,tgr3}, exploiting the GWTs emitted by BBHs.

Several recent papers \cite{paniPRD98,maggioPRD100,cardosoPRL116,cardosoPRD94,wangPRD97,zimmermanPRD96,abediPRD96,westerweckPRD97,nielsenPRD99,ricoPRD99,tsangPRD98,tsangPRD101} addressed the topic of \textit{exotic compact objects} (ECOs) \cite{cardosoECO}: possible compact objects (COs) alternative to the BHs predicted by Albert Einstein's GR theory.
Examples of ECOs include wormholes \cite{wormholes}, boson stars \cite{boson_star}, gravastar \cite{gravastar}, and fuzzballs \cite{fuzzball}.
These ECO models are characterized by different astrophysical properties, like their constituent ``matter'', but they all share one physical characteristic: Planck-scale modifications of the BH event horizon due to quantum effects\cite{wangPRD97} or the presence of a surface of different nature \cite{bh_horizon,cardosoPRD94}.
This feature would enable the emission of repeated GWTs occurring shortly after the BBH merger time, \textit{echoes} of the ECO remnant ringdown \cite{paniPRD98,maggioPRD100,wangPRD101}.

In this work, we report a systematic search for echo signals of generic morphology occurring after the merger-ringdown phase of BBH GWTs \cite{andreaphd}.
The detection performance of the method is demonstrated down to low signal-to-noise (SNR) ratios, and the results are practically independent of the echo signal morphology.
We also provide upper limits on the strain amplitude at earth of echoes for the loudest BBH GWTs in the LVK catalogs \cite{gwtc1,gwtc2,gwtc3}.
The search is based on the coherent Wave Burst pipeline (cWB) \cite{cwb2016,cwb2008,cwb2005,cwbSoftX,CWB} widely used to search for generic short duration GWTs by LVK \cite{allsky_o1,allsky_o2,allsky_o3}.

Section \ref{sec:echoes} provides a brief review of GW echo models and discusses the main characteristics of the predicted echo signals.
In section \ref{sec:method} we summarize the data analysis method focusing on novel features: in particular, on the search for weak post-merger-ringdown GWTs, the simulations with software signal injections and the construction of the confidence belt on echoes' $h_\mathrm{rss}$ \cite{sutton_arxiv} strain amplitude.
Section \ref{sec:results} reports the search results including detection performances, checks of robustness to different echo morphologies, and upper limits on the echoes' $h_\mathrm{rss}$.
A comparative discussion with respect to four, previously published, echo searches \cite{tsangPRD98,ricoPRD99,westerweckPRD97,abedifull} is reported in section \ref{sec:comparison}.
Conclusions are drawn in section \ref{sec:conclusions}.


\begin{figure}
    \centering
    \includegraphics[width=0.48\textwidth]{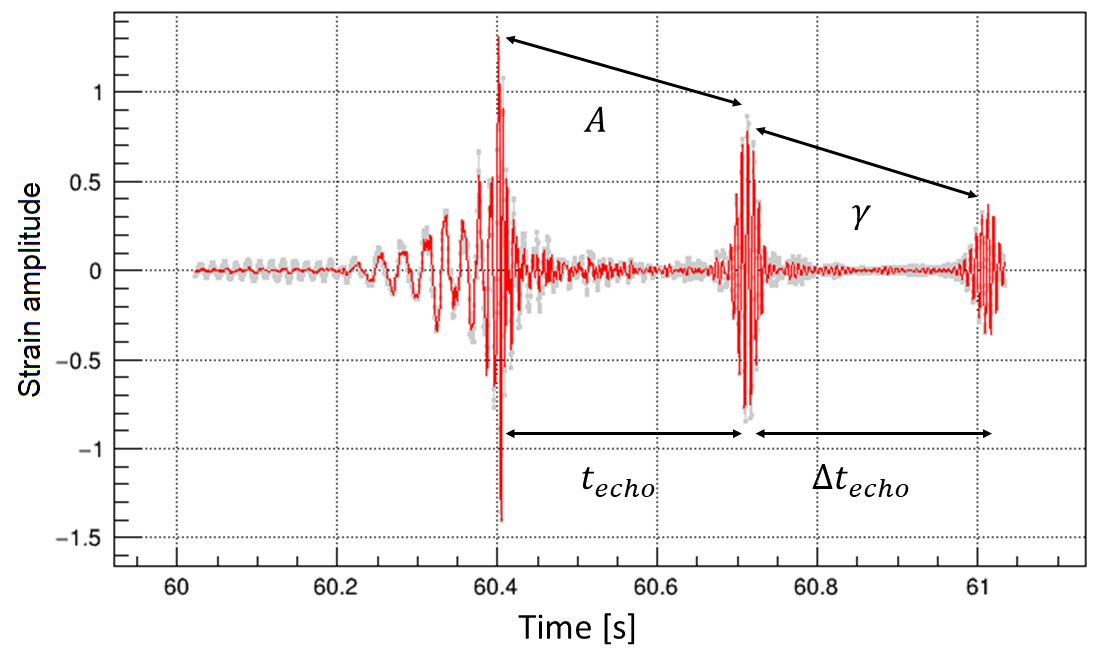}
    \caption{Simulated inspiral-merger-ringdown-echoes GW transient signal vs. time at one detector: the red line is the whitened reconstructed signal strain $h(t)$; the grey line is the whitened data.
    The CBC signal peaks at the merger time ($\SI{60.42}{\second}$) and is followed by echoes, whose main parameters are visualized.}
    \label{fig:echo_wf}
\end{figure}

\section{Gravitational Wave Echoes}
\label{sec:echoes}
Distinctive properties of an ECO remnant derives from the dynamics of its \textit{inner barrier}, interpreted as an \textit{effective surface}, located above the would-be BH event horizon (EH), $r_\mathrm{eh}$, \cite{cardosoPRD90,cardosoPRL116,cardosoPRD94} at a radius $r_\mathrm{in}$ from the object's center
\begin{equation}
	r_\mathrm{in} = r_\mathrm{eh} + \mathit{l} \, .
	\label{eq:rECO}
\end{equation}
Here, in eq.(\ref{eq:rECO}), $\mathit{l}$ is the \textit{length correction} to the would-be BH event horizon (EH), and it is theorised to be extremely small \cite{cardosoPRD94,wangPRD97,maggioPRD100}, of the order of the Planck length ($l_\mathrm{Planck} \sim 2 \cdot 10^{-35} \si{\metre}$).
The inner barrier couples to the outer one (i.e. angular-momentum potential barrier) acting as a sort of cavity \cite{zimmermanPRD96}.
If the remnant compact object of a CBC is not a GR BH, i.e. it is not fully absorbing, the merger-ringdown can trigger multiple outer and inner barrier excitations \cite{paniPRD98,maggioPRD100}.
This results in a train of pulses of outgoing GW radiation of decreasing energy called \textit{echoes} \cite{ferrariPRD62,maselliPRD96}.

Since the performance of our method is independent to details on the phase evolution of the signal, we can safely test it under the simplifying assumption of a non-rotating ECO remnant
\footnote{According to e.g. \cite{maggioPRD100}, echoes from a spinning remnant show a much larger variety of morphologies, including e.g. a redshift of the echo central frequency vs pulse order as well as different relative contributions of the $+$ and $\times$ GW polarization components at different pulses.
However, the sensitivity of our search is invariant when expressed in terms of the $h_{rss}$ component that is detectable by the GW observatory for a given GWT.
Of course, the interpretation of our measurement in terms of the source model will be different.}.
This assumption has also been adopted in many searches in the literature 
\cite{wangPRD97,zimmermanPRD96,abediPRD96,westerweckPRD97,nielsenPRD99,ricoPRD99,tsangPRD98}.
A complete description of a possible echo template for a spinning ECO remnant, as expected from a compact binary coalescence, is provided in \cite{maggioPRD100}.
The main parameters characterizing the models of echoes are \cite{westerweckPRD97,nielsenPRD99} (see also figure \ref{fig:echo_wf}):
\begin{itemize}
    \item{$\Delta t_\mathrm{echo}$ : the time separation between subsequent pulses as measured by a distant observer.
    It corresponds to the round-trip travel time of the space-time perturbation between the inner and outer barriers \cite{cardosoPRD94};}
    \item{$t_\mathrm{echo}$ : the delay of the first echo pulse from the coalescence time of the binary.
    In general $t_\mathrm{echo} \sim \Delta t_\mathrm{echo}$ apart from small effects related to the strong non-linearity close to the merger time;}
    \item{$\gamma$ : the attenuation per round-trip in terms of the GW amplitude ratio between subsequent echo pulses ($0 < \gamma < 1$);}
    \item{$A$ : amplitude ratio between the amplitude of the first echo and the one at merger time ($0 \leq A < 1$).}
\end{itemize}

Following \cite{cardosoPRL116,cardosoPRD94}, the theoretical prediction for $\Delta t_\mathrm{echo}$ is clearly related to the space-time geometry outside the ECO remnant:
    \begin{equation}
    \Delta t_\mathrm{echo} \sim 2 \int_{r_\mathrm{in}}^{r_\mathrm{out}}\frac{1}{\sqrt{F(r)B(r)}}dr \,
    \label{eq:dt_echo_metric}
    \end{equation}
where in eq. \ref{eq:dt_echo_metric}, $F(r)$ and $1/B(r)$ 
\footnote{The space-time geometry outside an ECO remnant can be described with the $ds^2 = -F(r)dt^2+1/B(r)dr^2+r^2d\Omega^2$ metric.
Such a metric is used to generally describe a static CO with spherical symmetry and matter localised only in the region $r<r_{shell}$.
Following Birkhoff's theorem, in the region $r>r_{shell}$ the Schwarzschild metric holds: $F(r)=B(r)=\left(1-\frac{2GM}{c^2r}\right)$.}
are the coefficients functions for the time and radial component of the metric in a spherically symmetric system, $r_\mathrm{in}$ is the radius of the inner barrier and $r_\mathrm{out}$ the radius of the outer barrier.
Eq. \ref{eq:dt_echo_metric} takes into account the effects of gravitational redshift and spatial curvature on the emission of GW echoes.
The resulting approximate expression for the time separation is \cite{cardosoPRL116,cardosoPRD94}:
\begin{equation}
    \Delta t_\mathrm{echo} \approx 54\left(\frac{n}{4}\right)M_{30}\left[1-0.001\ln\left(\frac{l/l_\mathrm{Planck}}{M_{30}}\right)\right] \, \si{\milli\second} \, .
    \label{eq:dt_echo_M30}
\end{equation}
Here, $n$ is a parameter of the order of the unity that takes into account the structure of the ECO nature \cite{cardosoPRD94,abediPRD96}, $M_{30} \equiv M/30M_{\odot}$ with $M$ standing for the final mass of the remnant.
Therefore, a measurement of $\Delta t_\mathrm{echo}$ would provide information over the theorized nature of the ECO through the parameters $n$ and $\mathit{l}$, related to the compactness of the ECO \cite{paniPRD98}.
According to eq.(\ref{eq:dt_echo_M30}), typical values for echoes time separation are $\Delta t_\mathrm{echo}\in(30,400) \; \si{\milli\second}$ for BBH mergers whose total mass ranges in $\in(10,100) \; M_{\odot}$, like most of those detected during O1, O2 and O3 by the LV Collaborations \cite{gwtc1,gwtc2,gwtc3}.


\subsection{Signal proxy for echoes}
\label{subsec:echo_proxy}
Our detection algorithm does not make use of signal templates, and for testing its performance we can rely on loose signal proxies.
The template we selected to mimic echo signals $h_\mathrm{echo}(t)$ is a double sine-Gaussian (SGE) pulse $h_\mathrm{echo}(t) = h_\mathrm{SGE}(t) + \gamma\cdot h_\mathrm{SGE}(t-\Delta t_\mathrm{echo})$ with $h_\mathrm{SGE}(t)$ \cite{SGEwf}:
\begingroup
\addtolength{\jot}{1.5em}
\begin{align}
\begin{split}
    h_{+,\mathrm{SGE}}(t) &= h_0 \, \frac{1+\cos^2(\iota)}{2} \, e^{ -\frac{t^2}{\tau^2} } \, \sin(2 \pi f_0 t + \phi_0)\, \\
    h_{\times,\mathrm{SGE}}(t) &= h_0 \, \cos(\iota) \, e^{ -\frac{t^2}{\tau^2} } \, \cos(2 \pi f_0 t + \phi_0)\, ,
\end{split}
\label{eq:sge}
\end{align}
\endgroup
In eq.(\ref{eq:sge}), $h_0$ is the signal amplitude, $\iota$ is the inclination angle of the source, $\tau$ the half-time duration of the pulse, $f_0$ and $\phi_0$ its central frequency and phase respectively.
The values we select for these parameters are:
\begin{itemize}
    \item{$h_0$ is defined as $h_0 = A \cdot h_\mathrm{max}$, where $h_\mathrm{max}$ is the GW amplitude at the merger. 
    In our simulations, $A$ is randomly selected per each injection within a uniform distribution $0 < A < 1$ (see \ref{subsec:off-source_exp}).}
    \item{$\gamma = 0.5$ so that the second echo is contributing 1/3 of the injected SNR.
    This is an intermediate condition on the concentration of the signal in time and makes possible to study the reconstruction of a weaker echo, separately from the first.}
    \item{$\tau = \SI{20}{\milli\second}$ and $f_0 = \SI{140}{\hertz}$, are close to expectations for the typical mass range of BBH mergers in GWTC-3.}
    \item{$\phi_0 = 0$. This is not impacting the results since the search method is agnostic on the signal phase in each pulse.}
    \item{$t_\mathrm{echo} = \SI{300}{\milli\second}$ and $\Delta t_\mathrm{echo} = \SI{300}{\milli\second}$ are intermediate values for the investigated BBH mergers (see section \ref{subsec:final_setup}) according to eq. \ref{eq:dt_echo_M30}.}
    \item{$\iota$ is set equal to the one of the injected BBH signal.}
\end{itemize}
Furthermore, the sky location of the echo signal proxy is the same as the one of the BBH GWT.


\section{Search methods}
\label{sec:method}
This section describes the methods developed to search for generic GWTs after BBH mergers, such as echo signals.
The analysis is based on 
cWB methods and comprises Monte Carlo simulations to tune the search and interpret the results in terms of gravitational wave echoes.
We call this new analysis \textit{cWB echo signal (ES) search}.

 
\subsection{Coherent WaveBurst}
\label{subsec:cWB}
Coherent WaveBurst \cite{CWB,cwbSoftX} is a data analysis pipeline searching for generic GWT signals in the data from the LVK GW detectors network \cite{ligoweb,virgoweb,kagraweb}.
Designed to operate without a specific waveform model, cWB first identifies coincident excess power in the multi-resolution time-frequency (TF) representations of the detectors’ strain data \cite{cwb2005}.
Then, for the selected events, cWB reconstructs the source sky location and the signal waveform of each GW candidate by means of a constrained maximum likelihood method \cite{cwb2008}.

To be robust against the non-stationary detector noise cWB employs signal-independent vetoes, reducing the initial high rate of the excess power triggers.
The primary selection cut is on the network correlation coefficient $c_\mathrm{c}$ \cite{cwb2016}, defined as:
\begin{equation}
	c_\mathrm{c} = \frac{E_\mathrm{c}}{E_\mathrm{c} + E_\mathrm{null}} \, ,
	\label{eq:cc}
\end{equation}
which is informative on the coherence of a signal among the detectors of the network.
Here, $E_\mathrm{c}$ and $E_\mathrm{null}$ \cite{suttonPRD74,cwb2008,cwb2016} are the coherent and the null energy of the signal.
The algorithm also combines all the data streams into one coherent statistic $\eta_\mathrm{c}$ \cite{cwb2016}, which is used for ranking the detected events and is defined as:
\begin{equation}
	\eta_\mathrm{c} = \sqrt{\frac{c_\mathrm{c} \cdot E_\mathrm{c}}{N-1}} \, ,
	\label{eq:rho}
\end{equation}
with $N$ the number of detectors in the network.
Typically, for a GW signal $c_\mathrm{c} \sim 1$ while for instrumental glitches $c_\mathrm{c} \ll 1$.
By setting a threshold value on $c_\mathrm{c}$, it is possible to reconstruct events with a lower or higher probability of being genuine GW signals.

In the LVK analyses, different cWB searches are used depending on the target GWT.
In a previous work, cWB was used to investigate post-merger GW emission in a configuration more sensitive to the chirping morphology of the CBCs signals \cite{cwb_widerlook}.
Currently, the most general cWB search is the all-sky burst search \cite{allsky_o1,allsky_o2,allsky_o3}, with a proven ability to detect the broadest variety of GW signal morphologies.
Our search method is based on this cWB instance, the same version used in the LVK O3 analysis \cite{allsky_o3,cwbSoftX}, thus it is more agnostic than \cite{cwb_widerlook}.
The following subsections describe the peculiarities of cWB ES search.


\begin{figure}
    \centering
    \includegraphics[width=0.48\textwidth]{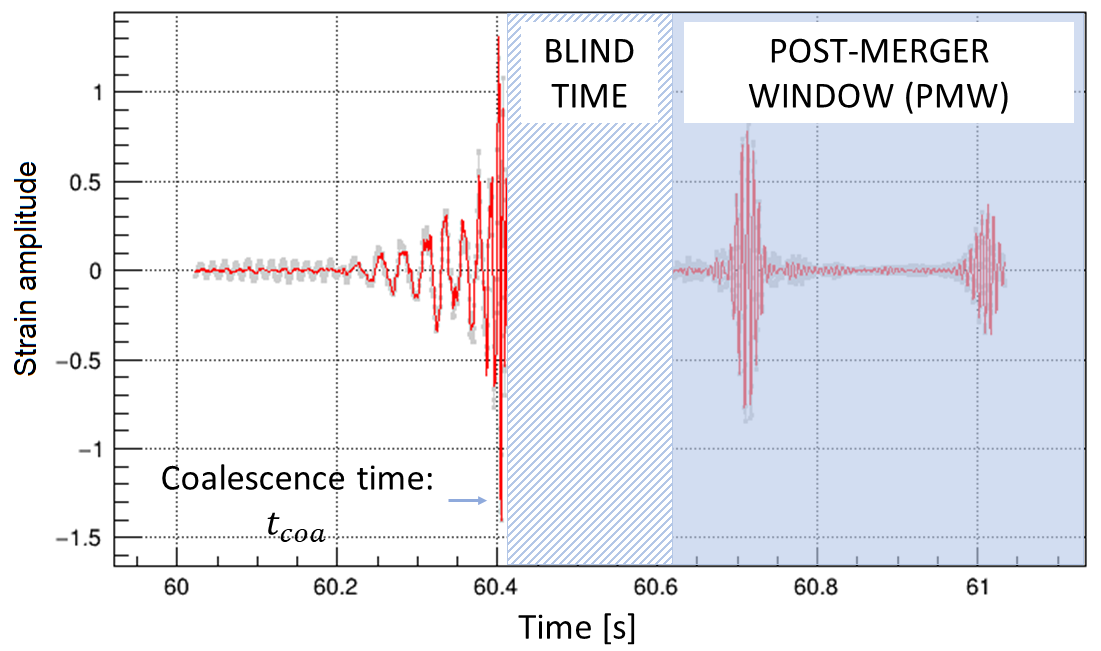}
    \caption{This plot shows the segmentation of the analyzed time following up an event (red line).
    The pale blue opaque area is representative of the blind time $\Delta t_\mathrm{blind}$, and the light blue transparent area after it highlights the post-merger window (PMW).}
    \label{fig:PMW_im}
\end{figure}

\subsection{Searching for echoes}
\label{subsec:pm}
Due to the expected nature of echoes, the cWB all-sky burst search is modified to select more TF pixels with a low energy content and scattered over a wider than usual time span (see appendix \ref{app:th_summary}).
Triggering and final selection thresholds are decreased, and to group different pulses (i.e. the BBH merger and the echo-like signals) into a single event, we increase the maximum time separation between disjoint clusters of pixels which define a single event.
Specifically, the $\eta_\mathrm{c}$ threshold is decreased from 5.0 to 3.5, and the $T_\mathrm{gap}$ parameter \cite{CWB} is increased up to $\SI{2}{\second}$.
Also, the whitening \cite{cwb_wht} of the data is performed using a TF map resolution which mitigates the leakage of the merger-ringdown signal of the remnant into the subsequent TF pixels.
Indeed, while the cWB all-sky burst search performs the whitening in the TF map with the best frequency resolution, typically $\Delta f = \SI{1}{\hertz}$ and $\Delta t = \SI{0.5}{\second}$, here we adopt a better time resolution, using pixels with a time width of $\Delta t = \SI{0.125}{\second}$ and $\Delta f = \SI{4}{\hertz}$.

The search uses the BBH GWT as a trigger and focuses on a user-defined post-merger time interval, called \textit{post-merger window} (PMW), see figure \ref{fig:PMW_im}.
The PMW starts at time $t_\mathrm{start}^{\scaleto{\mathrm{PMW}}{4pt}}$, defined as
\begin{equation}
    t_\mathrm{start}^{{\scaleto{\mathrm{PMW}}{4pt}}} = t_\mathrm{coa} + \Delta t_\mathrm{blind} \, ,
    \label{eq:tstartPMW}
\end{equation}
where $t_\mathrm{coa}$ is the coalescence time of the BBH system and $\Delta t_\mathrm{blind}$ a user-defined blind time.
The blind time's purpose is to mask the ringdown of the BBH signal, and its impact on the analysis will be discussed in \ref{subsec:final_setup}.
Limiting the ES search to a PMW allows to limit the noise contribution in the post-merger without penalising the capability to detect possible echo signals.
We adopted two choices of PMW: 
\begin{itemize}
	\item{for $t_\mathrm{echo} \leq \SI{200}{\milli\second}$ we use $\Delta t_\mathrm{blind} = \SI{50}{\milli\second}$ and $\Delta t^{\scaleto{\mathrm{PMW}}{4pt}} = \SI{300}{\milli\second}$;}
	\item{for $t_\mathrm{echo} > \SI{200}{\milli\second}$ we use $\Delta t_\mathrm{blind} = \SI{200}{\milli\second}$ and $\Delta t^{\scaleto{\mathrm{PMW}}{4pt}} = \SI{1}{\second}$.}
\end{itemize}
Such choices are suitable to include the first $\sim$1-4 echo pulses according to eq.(\ref{eq:dt_echo_M30}).

Within the PMW, the main statistical parameters we compute are the network correlation coefficient, $c_\mathrm{c}^{\scaleto{\mathrm{PMW}}{4pt}}$, analogous to $c_\mathrm{c}$ (see eq.(\ref{eq:cc})), and the network signal to noise ratio of the data, SNR$^{\scaleto{\mathrm{PMW}}{4pt}}$, defined as
\begin{equation}
    \text{SNR}^{\scaleto{\mathrm{PMW}}{4pt}} = \sqrt{\sum_{k}^{N}\sum_{j \in J} (x_\mathrm{k}[j])^2} \, ,
    \label{eq:stat}
\end{equation}
where $J$ is the set of the TF pixels corresponding to times inside $\Delta t^{\scaleto{\mathrm{PMW}}{4pt}}$, and $x_\mathrm{k}[j]$ are the whitened reconstructed data.

While cWB can work with arbitrary detectors networks, the ES search deployed here is run only over the two LIGO detectors network (H, Hanford, and L, Livingston \cite{adLIGO}).
The motivation is two-fold.
On one side, H and L detect most of the GWTs' SNR.
Moreover, under O2 and O3 conditions, the detection performance of minimally modeled searches for GWTs results more effective when restricted to the LIGO network of almost co-aligned detectors \cite{allsky_o3}.
This comes from a complex balance between background rejection capabilities against collection of GW information, as the target signal parameter space becomes higher dimensional when searching also Virgo \cite{adVIRGO} due to the need of taking into account both GW polarization components.


\subsection{Monte Carlo estimators}
\label{subsec:off-source_exp}
The ES search follows a two-track scheme: the background (BKG) analysis, and the signal (SIG) analysis.
Both analyses are off-source experiments, meaning that the data do not include the times corresponding to the detected GW signals.
The ES search is separately performed for each BBH GWT considered.

The \textbf{background (BKG) analysis} is used to estimate the noise statistics for the null hypothesis in the PMW.
We create a set of off-source software signal injections over the data stream using waveform templates of the specific BBH event under study.
These templates are randomly selected from the CBC waveform posterior samples \cite{GWOSC1,GWOSC2}, provided by the Parameter Estimation (PE) methods for the considered GW event (following the approximants used in \cite{gwtc1,gwtc2,gwtc3}).
The signals are injected widely separated, i.e. one each $\SI{600}{\second}$, to avoid systematic interferences in the analysis.

The \textbf{signal (SIG) analysis} enables the measurement of the sensitivity of the ES search to signals within the PMW.
The injected BBH GWTs are the same as the BKG analysis with, in addition, the injection of \textit{secondary signals} after each BBH merger according to the echo model of section \ref{subsec:echo_proxy}.
Different morphologies of secondary signals have been tested as well, see \ref{subsec:final_setup}.

\begin{figure}
    \centering
    \includegraphics[width=0.48\textwidth]{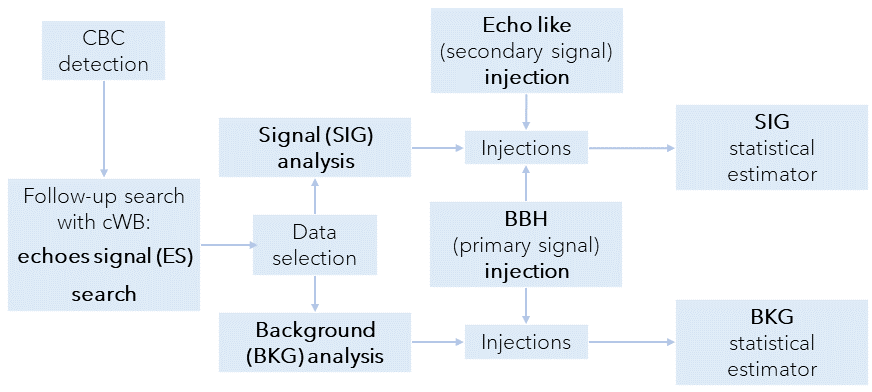}
    \caption{Flowchart of the echo signal (ES) search.
    Once the cWB all-sky burst search detects a BBH event, the cWB ES search can be run as a follow-up.
    The search runs two parallel studies on a common data selection: the background (BKG) and the signal (SIG), and computes all the statistical estimators described in section \ref{subsec:pm}.
    The BBH primary signal injections are randomly picked from the PE samples distribution for that event.}
    \label{fig:ES_scheme}
\end{figure}

This double simulation scheme is depicted in figure \ref{fig:ES_scheme}.
The data used for all studies are real data available at the GW open science center of the LVK collaboration, see \cite{GWOSC1,GWOSC2}.
These two analyses allow us to study the detection probability, DP, and the false alarm probability, FAP, as functions of the reconstructed SNR$^{\scaleto{\mathrm{PMW}}{4pt}}_\mathrm{rec}$.
Their definition is the following:
\begingroup
\addtolength{\jot}{1.5em}
\begin{align}
\begin{split}
	\text{DP} &= \frac{\text{EV}_{\text{SIG}}(\text{SNR}^{\scaleto{\mathrm{PMW}}{4pt}}_\mathrm{rec} \geq th_\mathrm{snr})}{\text{EV}} \\
	\text{FAP} &= \frac{\text{EV}_{\text{BKG}}(\text{SNR}^{\scaleto{\mathrm{PMW}}{4pt}}_\mathrm{rec} \geq th_\mathrm{snr})}{\text{EV}} \, ,
\end{split}
\label{eq:DE_FAP}
\end{align}
\endgroup
and here $\text{EV}_{\text{SIG}}$ and $\text{EV}_{\text{BKG}}$ are the number of detected events above threshold in the PMW from the SIG and BKG distributions, $\text{EV}$ is the total number of injected signals, and $th_\mathrm{snr}$ is the threshold value on SNR$^{\scaleto{\mathrm{PMW}}{4pt}}_\mathrm{rec}$.

\subsection{Tuning the analysis internal thresholds}
\label{subsec:tuning}
The cWB internal thresholds, described in \ref{subsec:pm} and listed in appendix \ref{app:th_summary}, are related to the energy content of a possible trigger, its energy per degree of freedom, and its coherence within the detectors' network.
On the contrary, they are agnostic to the signal morphology or spectral characterisation, so they allow to address a very wide range of different statistical noise conditions.

The tuning criteria of the internal thresholds are based on the receiver operating characteristics (ROC) curves, which are built from the DP and FAP measurements.
The chosen configuration of the analysis is the one that maximises the DP for low values of FAP, in the interval $0.5\% \leq$ FAP $\leq 5\%$.
This region corresponds to the events which possess low to medium SNR$^{\scaleto{\mathrm{PMW}}{4pt}}_\mathrm{rec}$, typically $4 \leq$ SNR$^{\scaleto{PMW}{4pt}}_\mathrm{rec}$ $\leq 8$.
The tuning has been extensively performed on the simulations related to the GW150914 event \cite{gw150914_det,gw150914_ob,gw150914_tgr}.
We have checked as well that the same setup is also providing the best results for a few GWTs from O3, including GW190521 \cite{PRD125GW190521}.

The list of tuned parameters is reported in appendix \ref{app:th_summary}.

\begin{figure}
    \centering
    \includegraphics[width=0.48\textwidth]{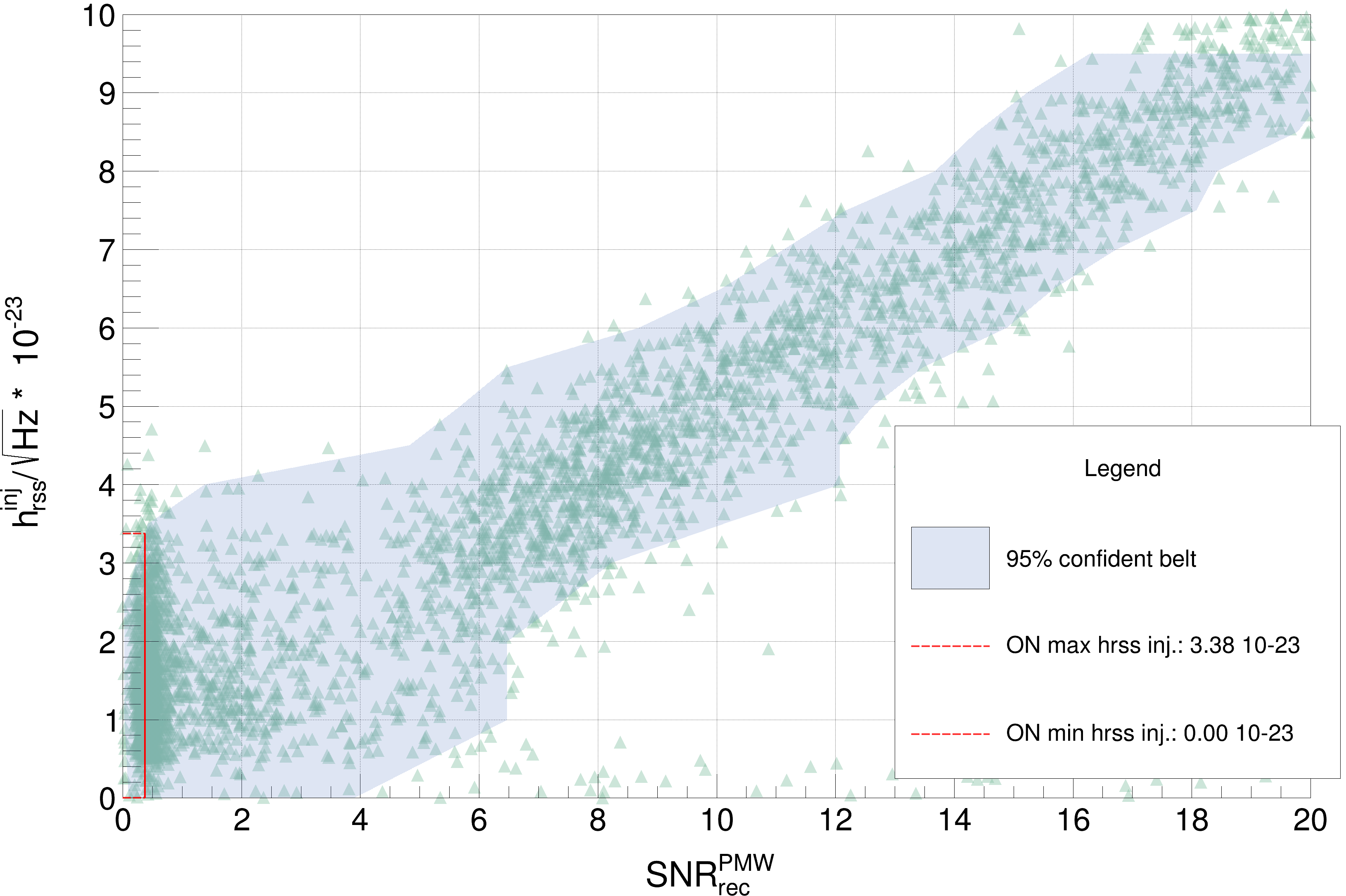}
    \caption{Confidence belt for the echo's injected amplitude $h_\mathrm{rss}$, $h_\mathrm{rss}^\mathrm{inj}$, vs the reconstructed SNR$^{\scaleto{\mathrm{PMW}}{4pt}}_{rec}$ in the PMW for GW150914.
    The blue region corresponds to $95\%$ coverage.
    The on-source $95\%$ confidence interval in terms of the $h_\mathrm{rss}$ is set by the intersection between the vertical line at the on-source value SNR$^{\scaleto{\mathrm{PMW}}{4pt}}_{\scaleto{\mathrm{ON}}{4pt}} \, \sim \, 0.38$ (red line) and the blue region.
    The y-axis values are in units of $10^{-23} / \sqrt{\si{\hertz}}$.
    The scatter plot shows SIG simulation data points but not BKG ones for readability.
    95$\%$ of the BKG samples has an SNR$^{\scaleto{\mathrm{PMW}}{4pt}}_{rec} \leq 4$.} 
    \label{fig:belt}
\end{figure}

\subsection{Inference of confidence intervals}
\label{subsec:pm_belt}
Searches for GWTs of generic morphologies, such as cWB ES, directly measure the \textit{energy} or \textit{integrated squared amplitude} of the candidate signal.
Here, results are presented in terms of $h_\mathrm{rss}$  \cite{sutton_arxiv} at earth
\begin{equation}
	h_\mathrm{rss} = \sqrt{\int_{t \in \mathrm{PMW}} ( \mid h_+(t) \mid^2 + \mid h_{\times}(t) \mid^2 ) dt}
	\label{eq:hrss}
\end{equation}
of signals consistent with the on-source data in the PMW.

The SIG simulations provide estimates of the conditional distributions of the recovered SNR$^{\scaleto{\mathrm{PMW}}{4pt}}_\mathrm{rec}$ as a function of $h_\mathrm{rss}^\mathrm{inj}$, the $h_\mathrm{rss}$ of the injected post-merger signal. 
We use the conditional distributions to build the confidence belts \cite{neymanRS_confidencebelt} on $h_\mathrm{rss}^\mathrm{inj}$, as shown in figure \ref{fig:belt}.
This is approximately achieved by introducing a binning in $h_\mathrm{rss}^\mathrm{inj}$ which ensures a minimum number of samples, hundreds per bin, and allows to target a confidence belt coverage of $95\%$.
The related cost is to perform specific SIG simulations with much higher statistics than those needed for estimating the DP. 
For the special case of $h_\mathrm{rss}^\mathrm{inj} = 0$, the null hypothesis, we exploit the full statistics of the BKG simulation.
The confidence belt is then used to set the $95\%$ confidence interval on the expected $h_\mathrm{rss}^\mathrm{inj}$ for signals inside the PMW that posses a SNR$^{\scaleto{\mathrm{PMW}}{4pt}}$ equal to the one measured on-source, SNR$^{\scaleto{\mathrm{PMW}}{4pt}}_{\scaleto{\mathrm{ON}}{4pt}}$.


\section{Results}
\label{sec:results}

\begin{table*}[t]
\centering
\footnotesize
\setlength{\tabcolsep}{6pt} 
\renewcommand{\arraystretch}{1.8} 
    \begin{tabular}{ l  c | c | c | c | c | c | c }
    \hline
    \hline
    \multicolumn{8}{ c }{List of analysed BBH events} \\
    \hline
    Run - GW name & App. & $t_\mathrm{coa}$ & SNR$_\mathrm{net}$ & $h_\mathrm{rss}^{50\%} \cdot \frac{10^{-23}}{\sqrt{\si{\hertz}}}$ & $t_\mathrm{echo}$ [$\si{\milli\second}$] & SNR$^{\scaleto{\mathrm{PMW}}{4pt}}_{\scaleto{\mathrm{ON}}{4pt}}$ & p-value$_\mathrm{ON}$ \\
    \hline
O1 - GW150914 		& 1 & 1126259462.421 & 24.4 & $ 2.79 \pm 0.02 $ & $ 227^{+12}_{-11} $ 	& $ 0.4 $ 	& $ 0.849 \pm 0.006 $ \\ 
O1 - GW151012 		& 1 & 1128678900.467 & 10.0 & $ 2.57 \pm 0.03 $ & $ 127^{+39}_{-14} $ 	& $ 0.1 $ 	& $ 0.54 \pm 0.01 $ \\ 
O1 - GW151226 		& 1 & 1135136350.668 & 13.1 & $ 2.70 \pm 0.03 $ & $ 73^{+23}_{-5} $ 	& $ 0.2 $ 	& $ 0.8 \pm 0.1 $ \\ 

O2 - GW170104 		& 1 & 1167559936.619 & 13.0 & $ 2.52 \pm 0.01 $ & $ 176^{+18}_{-14} $ 	& $ 0.4 $ 	& $ 0.673 \pm 0.006 $ \\ 
O2 - GW170608 		& 1 & 1180922494.501 & 14.9 & $ 2.63 \pm 0.01 $ & $ 63^{+12}_{-2} $ 	& $ 2.2 $ 	& $ 0.034 \pm 0.003 $ \\ 
O2 - GW170729 		& 1 & 1185389807.346 & 10.2 & $ 2.53 \pm 0.01 $ & $ 287^{+53}_{-37} $ 	& $ 2.5 $ 	& $ 0.043 \pm 0.003 $ \\ 
O2 - GW170809 		& 1 & 1186302519.758 & 12.4 & $ 2.40 \pm 0.02 $ & $ 203^{+19}_{-14} $ 	& $ \leq 0.1 $ 	& $ 0.56 \pm 0.02 $ \\ 
O2 - GW170814 		& 1 & 1186741861.533 & 15.9 & $ 2.51 \pm 0.02 $ & $ 191^{+12}_{-9} $ 	& $ 1.5 $ 	& $ 0.450 \pm 0.007 $ \\ 
O2 - GW170823 		& 1 & 1187529256.501 & 11.5 & $ 2.50 \pm 0.02 $ & $ 236^{+36}_{-27} $ 	& $ \leq 0.1 $ 	& $ 0.835 \pm 0.006 $ \\ 

O3a - GW190408\_181802 	& 2 & 1238782700.279 & 14.7 & $ 1.82 \pm 0.01 $ & $ 147^{+14}_{-10} $ 	& $ 1.1 $ 	& $ 0.155 \pm 0.005 $ \\ 
O3a - GW190412 		& 2 & 1239082262.165 & 18.9 & $ 1.82 \pm 0.01 $ & $ 134^{+14}_{-14} $ 	& $ 1.2 $ 	& $ 0.273 \pm 0.007 $ \\ 
O3a - GW190512\_180714 	& 2 & 1241719652.435 & 12.3 & $ 1.69 \pm 0.02 $ & $ 123^{+14}_{-12} $ 	& $ \leq 0.1 $ 	& $ 0.55 \pm 0.02 $ \\ 
O3a - GW190513\_205428 	& 2 & 1241816086.800 & 12.3 & $ 1.83 \pm 0.01 $ & $ 185^{+29}_{-21} $ 	& $ 0.2 $ 	& $ 0.882 \pm 0.004 $ \\ 
O3a - GW190517\_055101 	& 2 & 1242107479.848 & 10.2 & $ 1.80 \pm 0.02 $ & $ 213^{+33}_{-32} $ 	& $ \leq 0.1 $ 	& $ 0.52 \pm 0.01 $ \\ 
O3a - GW190519\_153544 	& 2 & 1242315362.418 & 12.0 & $ 1.84 \pm 0.01 $ & $ 365^{+45}_{-50} $ 	& $ 0.4 $ 	& $ 0.302 \pm 0.006 $ \\ 
O3a - GW190521 		& 2 & 1242442967.471 & 14.4 & $ 1.74 \pm 0.01 $ & $ 568^{+133}_{-81} $ 	& $ 0.2 $ 	& $ 0.858 \pm 0.004 $ \\ 
O3a - GW190521\_074359 	& 2 & 1242459857.456 & 24.4 & $ 1.73 \pm 0.01 $ & $ 256^{+23}_{-16} $ 	& $ 0.5 $ 	& $ 0.218 \pm 0.006 $ \\ 
O3a - GW190602\_175927 	& 2 & 1243533585.093 & 12.1 & $ 1.98 \pm 0.04 $ & $ 402^{+64}_{-54} $ 	& $ 0.2 $ 	& $ 0.838 \pm 0.005 $ \\ 
O3a - GW190701\_203306 	& 2 & 1246048404.578 & 11.6 & $ 1.84 \pm 0.01 $ & $ 326^{+41}_{-32} $ 	& $ 6.4 $ 	& $ 0.0019 \pm 0.0007 $ \\ 
O3a - GW190706\_222641 	& 2 & 1246487219.361 & 12.3 & $ 1.82 \pm 0.01 $ & $ 358^{+66}_{-49} $ 	& $ 0.3 $ 	& $ 0.100 \pm 0.004 $ \\ 
O3a - GW190814 		& 2 & 1249852257.009 & 22.2 & $ 1.82 \pm 0.01 $ & $ 91^{+4}_{-3} $ 	& $ 0.5 $ 	& $ 0.718 \pm 0.005 $ \\ 
O3a - GW190828\_063405 	& 2 & 1251009263.781 & 16.0 & $ 1.82 \pm 0.01 $ & $ 197^{+26}_{-15} $ 	& $ 1.0 $ 	& $ 0.474 \pm 0.006 $ \\ 
O3a - GW190915\_235702 	& 1 & 1252627040.693 & 13.1 & $ 1.88 \pm 0.02 $ & $ 205^{+26}_{-22} $ 	& $ 0.2 $ 	& $ 0.720 \pm 0.008 $ \\ 
O3a - GW190929\_012149 	& 1 & 1253755327.505 & 9.9 & $ 1.86 \pm 0.02 $ & $ 367^{+122}_{-92} $ 	& $ 0.1 $ 	& $ 0.860 \pm 0.005 $ \\ 

O3b - GW191109\_010717 	& 2 & 1257296855.783 & 17.3 & $ 1.85 \pm 0.01 $ & $ 387^{+65}_{-54} $ 	& $ 0.1 $ 	& $ 0.715 \pm 0.008 $ \\ 
O3b - GW191204\_171526 	& 2 & 1259514944.087 & 17.5 & $ 2.05 \pm 0.05 $ & $ 68^{+6}_{-4} $ 	& $ 0.3 $ 	& $ 0.36 \pm 0.02 $ \\ 
O3b - GW191215\_223052 	& 2 & 1260484270.995 & 11.2 & $ 1.69 \pm 0.02 $ & $ 148^{+18}_{-15} $ 	& $ 0.2 $ 	& $ 0.50 \pm 0.02 $ \\ 
O3b - GW191222\_033537 	& 2 & 1261020955.347 & 12.5 & $ 1.82 \pm 0.01 $ & $ 272^{+55}_{-36} $ 	& $ \leq 0.1 $ 	& $ 0.771 \pm 0.007 $ \\ 
O3b - GW191230\_180458 	& 2 & 1261764316.898 & 14.4 & $ 1.77 \pm 0.08 $ & $ 296^{+61}_{-40} $ 	& $ \leq 0.1 $ 	& $ 0.31 \pm 0.04 $ \\ 
O3b - GW200219\_094415 	& 2 & 1266140673.095 & 10.7 & $ 1.75 \pm 0.07 $ & $ 224^{+42}_{-28} $ 	& $ \leq 0.1 $ 	& $ 0.28 \pm 0.05 $ \\ 
O3b - GW200224\_222234 	& 2 & 1266618172.381 & 20.0 & $ 1.74 \pm 0.01 $ & $ 247^{+24}_{-17} $ 	& $ 7.4 $ 	& $ 0.0017 \pm 0.0005 $ \\ 
O3b - GW200225\_060421 	& 2 & 1266645879.413 & 12.5 & $ 1.89 \pm 0.04 $ & $ 115^{+13}_{-10} $ 	& $ 0.7 $ 	& $ 0.21 \pm 0.02 $ \\ 
O3b - GW200311\_115853 	& 2 & 1267963151.380 & 17.8 & $ 1.80 \pm 0.01 $ & $ 212^{+17}_{-14} $ 	& $ 0.3 $ 	& $ 0.561 \pm 0.006 $ \\ 
    \hline
    \hline
    \end{tabular}
    \caption{The analyzed BBH GWTs with SNR$_\mathrm{net} \geq 10$ in the all-sky cWB search \cite{allsky_o1,allsky_o2,allsky_o3}, listed in chronological order.
    Columns report the observing run; GWT name; waveform model used in BKG and SIG simulations (1 = IMRPhenomPv2 approximant, 2 = SEOBNRv4PHM approximant, as the one used in \cite{gwtc1,gwtc2,gwtc3}); GPS coalescence time $t_\mathrm{coa}$ to $\si{\milli\second}$ resolution from the PE waveform posterior information available through GWOSC \cite{GWOSC1,GWOSC2}; network SNR as reconstructed by the cWB all-sky search for burst GWTs; $h_\mathrm{rss}$ in PMW that ensures DP=$50\%$ at FAP=$5\%$; predicted $t_\mathrm{echo}$ (following eq.(\ref{eq:dt_echo_M30}) and assuming n=8, wormhole, and $\mathit{l}/l_\mathrm{Planck}=2$); on-source SNR inside the PMW, SNR$^{\scaleto{\mathrm{PMW}}{4pt}}_{\scaleto{\mathrm{ON}}{4pt}}$; p-value of SNR$^{\scaleto{\mathrm{PMW}}{4pt}}_{\scaleto{\mathrm{ON}}{4pt}}$ with statistical uncertainties.
    The analyzed PMW is [50,350]$\si{\milli\second}$ for GWTs with $t_\mathrm{echo} \leq \SI{200}{\milli\second}$, [0.2,1.2]$\si{\second}$ otherwise.}
\label{tab:event_list}
\end{table*}

Using the search tuning described in \ref{subsec:tuning}, we investigated a sub-set of 33 BBH events from the BBH detections from LVK collaboration \cite{gwtc1,gwtc2,gwtc3}.
The subset comprises all the BBH events that possess a network SNR greater than 10 in the cWB search for generic GWTs \cite{allsky_o1,allsky_o2,allsky_o3} \footnote{The network SNR recovered by cWB is consistent to the one recovered by template searches for these loud BBH events}.
The selection is motivated by the reasonable expectation that the signal amplitude of echoes is such that $A \ll 1$, since no signals with amplitude comparable to that of the merger have been observed after the ringdown phase of any BBH GW emission.
The list of investigated BBH events and related main results is given in table \ref{tab:event_list}.


\begin{figure}
    \centering
    \includegraphics[width=0.48\textwidth]{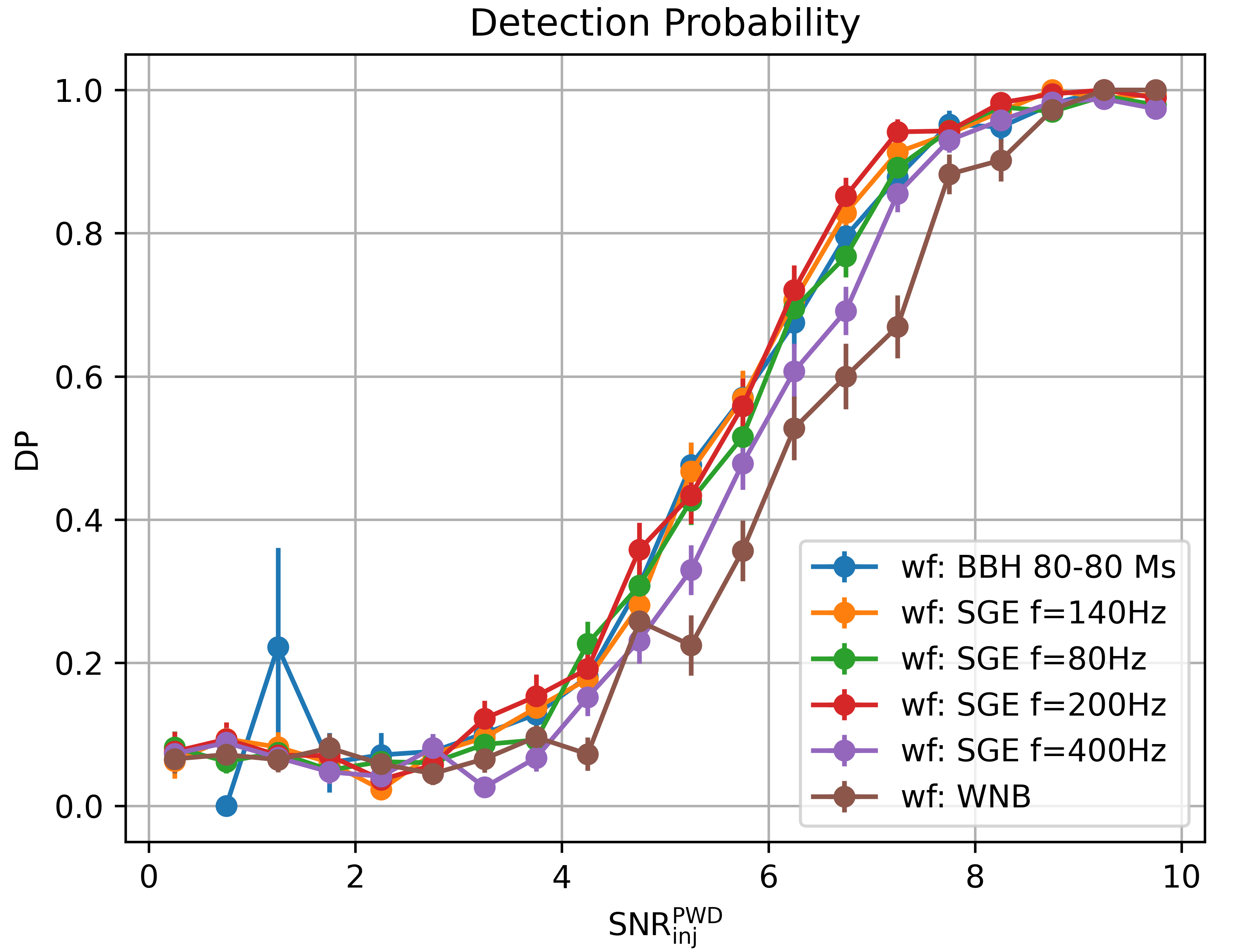}
    \caption{Detection probability (DP) at FAP = $5\%$ as a function of SNR$^{\scaleto{\mathrm{PMW}}{4pt}}_{\mathrm{inj}}$ for
    different morphologies of simulated post-ringdown signals: a high mass ($80-80 M_{\odot}$) BBH coalescence (blue), trains of two elliptically polarised sine-Gaussian pulses as described in \ref{subsec:echo_proxy} with different central frequencies $f_0 = 80, 140, 200, 400 \, \si{\hertz}$ (orange, green, red, and violet respectively) and a single pulse of white noise, WNB, of duration $\sim \SI{0.02}{\second}$, central frequency $\SI{150}{\hertz}$ and bandwidth $\SI{100}{\hertz}$ (brown). These results refer to GW150914.}
    \label{fig:DEvsECHO}
\end{figure}

\subsection{Robustness of cWB ES search}
\label{subsec:final_setup}
The BKG simulations show that the statistical properties of the noise background are weakly related to the choice of $\Delta t_\mathrm{blind}$ within the range $(0.05,0.4)\, \si{\second}$. 
Therefore, any $\Delta t_\mathrm{blind}$ in this range can be freely selected for the cWB ES search.
Instead, the noise level starts to increase as $\Delta t_\mathrm{blind}$ gets shorter due to some residual leakage from the primary BBH GWT signal into the PMW. 
The duration of the PMW window, $\Delta t^{\scaleto{PMW}{4pt}}$, affects as expected the mean SNR$^{\scaleto{\mathrm{PMW}}{4pt}}$ from the BKG analysis, the longer $\Delta t^{\scaleto{PMW}{4pt}}$ the larger the noise in the PMW.

We also tested the robustness of the cWB ES search against variations of the injected secondary signals in SIG analyses, see section \ref{subsec:off-source_exp}, for a few BBH GWT cases.
By changing the delay time $t_\mathrm{eco}$ and time separation $\Delta t_\mathrm{echo}$ of the two pulses of the signal proxy defined in section \ref{subsec:echo_proxy}, the detection probability at FAP=$5\%$ results unaffected as long as both pulses occur inside the analyzed time window, PMW.
Therefore, the off-source results reported in this work can be considered valid as long as $\Delta t^{\scaleto{PMW}{4pt}}$ and $\Delta t_\mathrm{blind}$ are included in the tested ranges,  [0.2.1.2]$\si{\second}$ and [0.05,0.2]$\si{\second}$ respectively, regardless of the choice $t_\mathrm{eco} = \Delta t_\mathrm{echo} = 0.3\,\si{\second}$ which we adopted in the SIG analyses of all BBH GWTs.

Moreover, we checked the sensitivity of the cWB ES search to widely different morphologies of post-ringdown signals, by performing additional SIG analyses.
Figure \ref{fig:DEvsECHO} shows the DP at FAP = $5\%$ as a function of the injected SNR$^{\scaleto{PMW}{4pt}}$ for different central frequencies of the SGE echo signal proxy (see section \ref{subsec:echo_proxy}), for a single pulse made by a BBH merger waveform and for a single burst of white noise.
The resulting performances are almost identical within uncertainties, which is an expected outcome due to the general nature of the cWB search (see section \ref{subsec:cWB}).
The slight decrease in performances when injecting white noise burst (WNB) signals in the PMW is mostly related to their wider frequency band.

\begin{figure}
    \centering
    \includegraphics[width=0.48\textwidth]{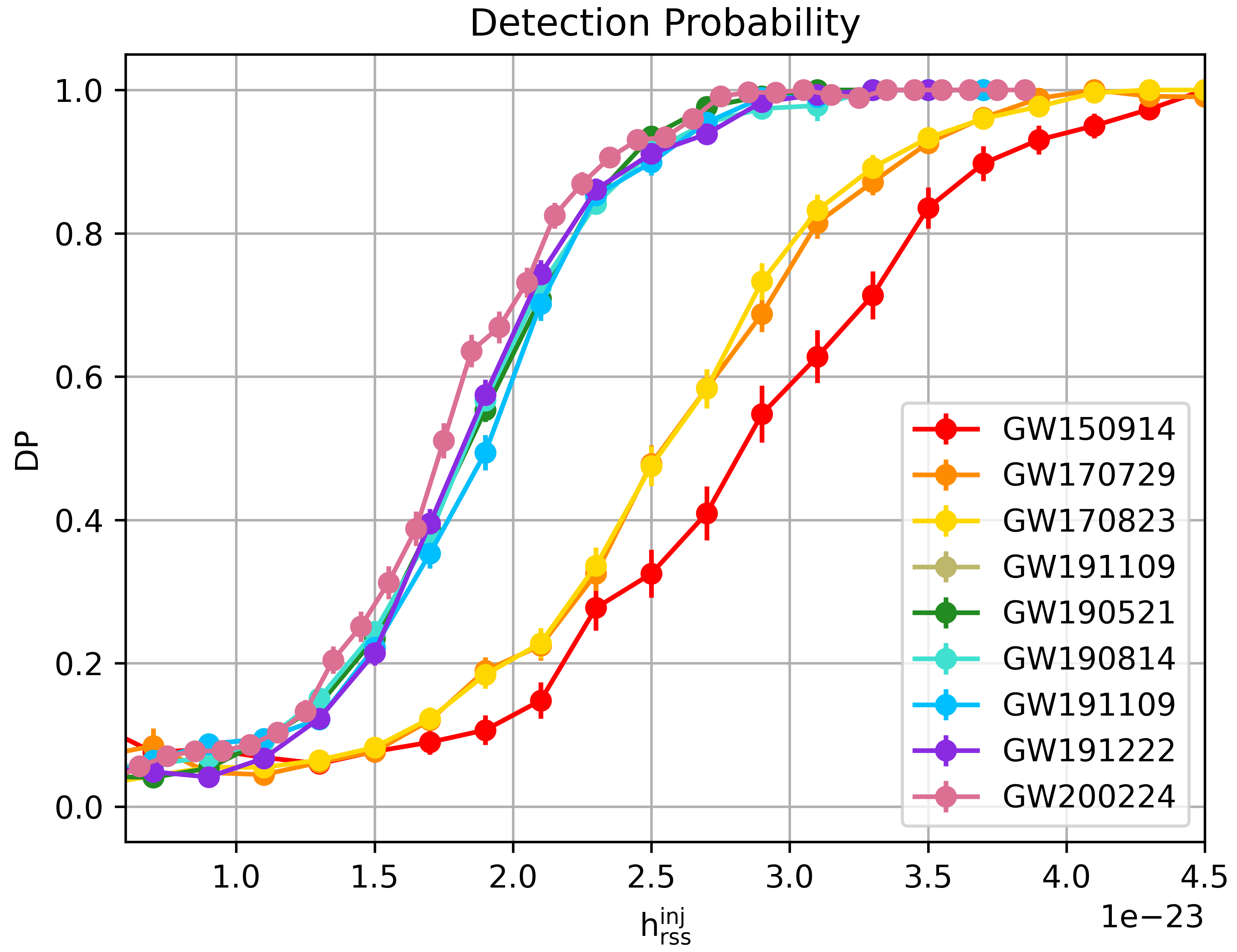}
    \caption{Plot of the detection probability as a function of $h_\mathrm{rss}^{\mathrm{inj}}$ of the echo signal proxy for a selection of BBH GWTs from each observing run of the LVK (O1, O2, and O3).
    The $h_\mathrm{rss}^{\mathrm{inj}}$ units are $10^{-23} \, \cdot \, \sqrt{\si{\hertz}}$.
    The temporal sensitivity improvement achieved is visible by the shift of the curves to lower $h_\mathrm{rss}^{\mathrm{inj}}$.}
    \label{fig:eff}
\end{figure}


\subsection{Detection probability}
\label{subsec:det-eff}
We discuss here the detection probability measurements for the echo signal proxy described in section \ref{subsec:echo_proxy}, with the requirement of FAP = $5\%$.
Figure \ref{fig:eff} shows the DP as a function of the $h_\mathrm{rss}$ injected inside the PMW for a subset of GWTs from the three LVK observing runs (O1, O2, O3).
The visible improvement towards smaller $h_\mathrm{rss}$ comes from the temporal enhancement of the detectors' sensitivities.
Between O1 and O2 observing runs, the typical  $h_\mathrm{rss}$ at $50\%$ DP decreases from $\sim 2.7 \cdot 10^{-23}/\sqrt{\si{\hertz}}$ to $\sim 2.5 \cdot 10^{-23}/\sqrt{\si{\hertz}}$.
A more significant decrease in $h_\mathrm{rss}$ at $50\%$ DP can be seen from O2 to O3, from average values of $\sim 2.5 \cdot 10^{-23}/\sqrt{\si{\hertz}}$ to $\sim 1.8 \cdot 10^{-23}/\sqrt{\si{\hertz}}$, corresponding to an improvement of about 28$\%$.
Column 6 of table \ref{tab:event_list} reports the resulting $h_\mathrm{rss}$ values which ensure $50\%$ DP with FAP $5\%$ for all the studied GWTs.

The coherent WaveBurst ES search explores a significantly lower range of $h_\mathrm{rss}$ values with respect to the cWB all-sky search for short-duration bursts \cite{allsky_o3}.
For the latter, the best results in terms of $h_\mathrm{rss}$ values at DP=$50\%$ among the tested signal morphologies has been achieved in O3 for a single pulse SGE, $Q=100$, $f_0=\SI{235}{\hertz}$, reaching $h_\mathrm{rss} = 8 \cdot 10^{-23}/\sqrt{\si{\hertz}}$ at a FAR of one per 100 years. 
Here instead, with a more dispersed signal, the double pulse SGE, Q = 8.8, $f_0=\SI{140}{\hertz}$, the average $h_\mathrm{rss}$ values at DP=$50\%$ in O3 reaches $\sim 1.9 \cdot 10^{-23}/\sqrt{\si{\hertz}}$, but at a much higher FAR of 2 per year, estimated by multiplying the FAP by the rate of the investigated BBH GWTs.


\subsection{On-source p-value}
\label{subsec:p-value}
\begin{figure}
    \centering
    \includegraphics[width=0.48\textwidth]{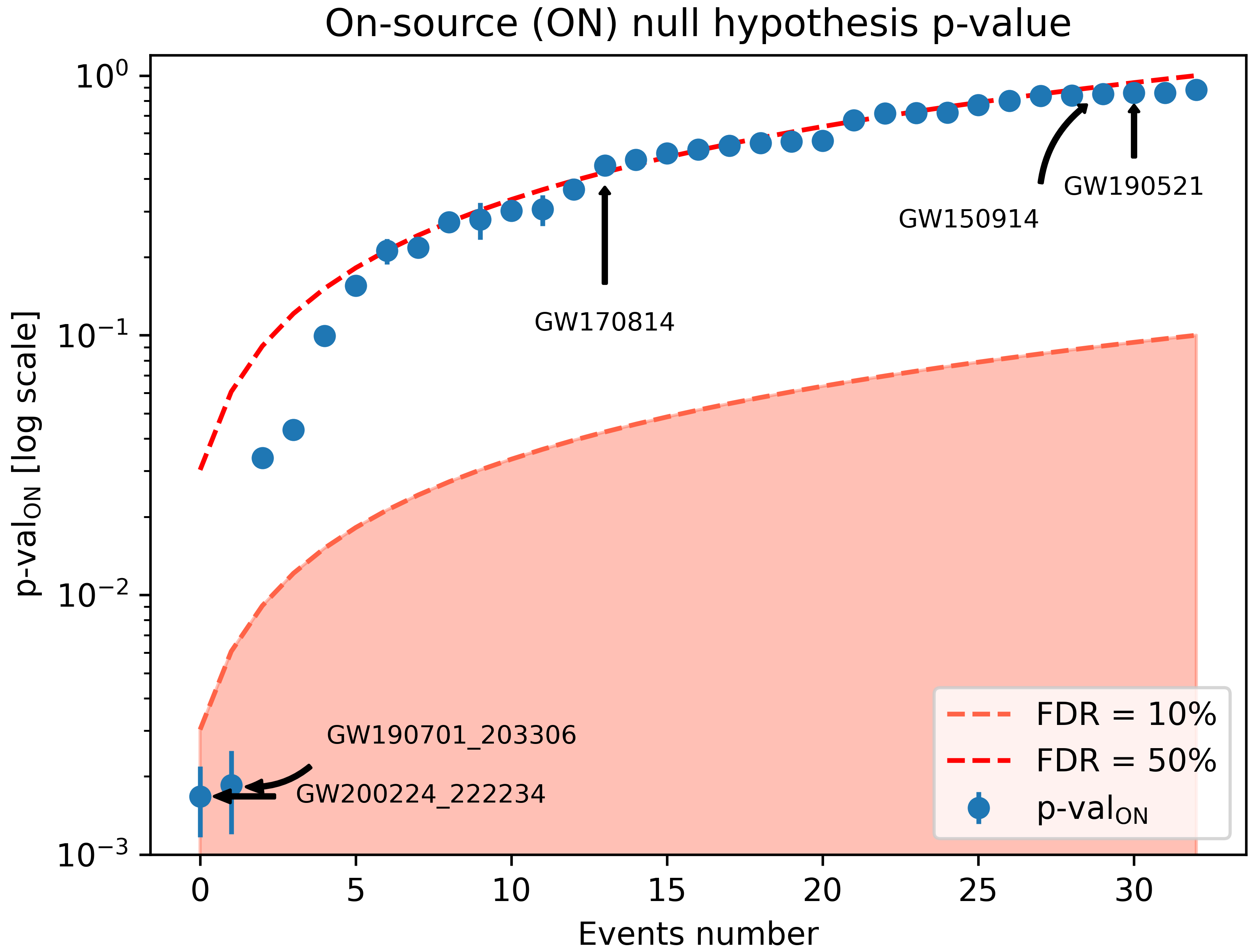}
    \caption{The plot shows the ordered p-values of SNR$^{\scaleto{\mathrm{PMW}}{4pt}}_{\scaleto{\mathrm{ON}}{4pt}}$ for the null hypothesis as measured by BKG analyses.
    P-values and their statistical uncertainties are represented by the blue dots.
    The \textbf{red dashed line} corresponds to the expected values for the null hypothesis (FDR$=50\%$).
    The \textbf{orange dashed line} and \textbf{orange filled area} corresponds to a FDR =$10\%$, and to the region FDR $<10\%$ respectively.
    This area is used to select echo candidates.}
    \label{fig:pvalue}
\end{figure}

The on-source (ON) data for each BBH GWT is analyzed using the same configuration of the cWB ES search of the SIG and BKG analyses (see sec. \ref{subsec:off-source_exp}).
By comparing the ON results with their BKG distributions we can estimate the p-value of SNR$^{\scaleto{\mathrm{PMW}}{4pt}}_{\scaleto{\mathrm{ON}}{4pt}}$ per each BBH GWT:
\begin{equation}
    \text{p-value}_{\text{ON}} = \frac{\text{EV}_{\text{BKG}}(\text{SNR}^{\mathrm{PMW}}_\mathrm{rec} \geq \text{SNR}^{\scaleto{\mathrm{PMW}}{4pt}}_{\scaleto{\mathrm{ON}}{4pt}})}{\text{EV}} \, ,
    \label{eq:pvalueON}
\end{equation}
where SNR$^{\scaleto{\mathrm{PMW}}{4pt}}_{\scaleto{\mathrm{ON}}{4pt}}$ is the on-source reconstructed SNR inside the PMW, EV is the total number of BKG instances and $\text{EV}_{\text{BKG}}$ is the number of BKG instances with SNR$^\mathrm{PMW}_\mathrm{rec}$ above the ON value. 
A low p-value points to SNR$^{\scaleto{\mathrm{PMW}}{4pt}}_{\scaleto{\mathrm{ON}}{4pt}}$ on the high-energy tail of the SNR$^\mathrm{PMW}_\mathrm{rec}$  distribution for the null hypothesis.
Columns 8 and 9 of table \ref{tab:event_list} list the SNR$^\mathrm{PMW}_\mathrm{ON}$ and p-value$_{\text{ON}}$ per each BBH GWT.
Figure \ref{fig:pvalue} reports the p-value for each investigated GWT, ranked from the lowest to the highest.
These estimates are based on the BKG analyses performed over approximately one calendar month of data around each BBH GWT.
We set an a priori threshold on the false discovery rate \cite{millerAJ122}, FDR $<\,0.1$,  to select the p-values hinting at a rejection of the null hypothesis. 
These cases are then the object of deeper follow-up studies.

Two GW events, GW190701 and GW200224, show an interesting SNR$^{\scaleto{\mathrm{PMW}}{4pt}}_{\scaleto{\mathrm{ON}}{4pt}}$ and their p-values pass the a priori FDR threshold.
In both cases, the morphological information of the outliers reconstructed inside the PMW (see appendix \ref{app:special_ev}) points to a dominant contribution by known instrumental disturbances in the frequency range $(16,40) \, \si{\hertz}$ \cite{caberoCQG32Hz,DQo1}.
These noise disturbances are known to often occur as a train of more pulses with a quasi-regular time separation.
This feature is especially evident in our analysis of GW200224 (see appendix \ref{app:gw200224}) and can affect our p-values estimates, since it violates the assumption of uniformly random occurrence times and of independence of each noise pulse.
Therefore, one can expect, at the very least, an underestimation of the uncertainties of our p-values.

We checked for systematic errors in the p-values of GW190701 and GW200224 by changing the off-source injection times of the BBH GWTs inside the BKG analysis.
In particular, we repeated the BKG analysis using only $\SI{4096}{\second}$ of data around the GWT time.
The new \textit{local} p-value estimates are
\begingroup
\addtolength{\jot}{1.5em}
\begin{align}
    &\text{GW190701:} \quad  \text{p-value}_{\text{ON}}^{\mathrm{local}} = 0.004 \pm 0.002 \\
    &\text{GW200224:} \quad  \text{p-value}_{\text{ON}}^{\mathrm{local}} = 0.007 \pm 0.002 \, ,
    \label{eq:p-value4096}
\end{align}
\endgroup
also reported in figure \ref{fig:pvalue_update}, in appendix \ref{app:special_ev}.
In the case of GW200224, the discrepancy between the estimates points to large systematic effects, including a significant bias of the p-value, which weakens its initial statistical significance.
As for GW190701, the local p-value estimate is also higher than the initial one, though it may still be compatible within the stated statistical uncertainties.

Further statistical checks and more morphological tests on GW190701 and GW 200224 are reported in appendix \ref{app:special_ev}.
Among these checks, the most important observation is that the reconstructed frequency spectrum for both the candidates does not match any expectation from echo models \cite{wangPRD97}, so these outliers cannot be considered plausible candidates for echoes.
We conclude that these two outliers are not suitable candidates for echo signals and are very likely instrumental disturbances.

For all the other GWTs, our p-value estimates occur well above our FDR threshold of attention, and their distribution is well described by the empirical BKG model.
Therefore, our work does not reject the null hypothesis, confirming what was previously reported by different search methods:
\begin{itemize}
    \item{the generic echo search of \cite{tsangPRD98}, which estimated p-values in the post-ringdown of the GWTs detected in observing runs O1 and O2 \cite{tsangPRD101} and in O3b \cite{tgr3};}
    \item{the template-based searches \cite{tgr1,tgr2} on O1, O2 data, and \cite{ricoPRD99,westerweckPRD97}, which provided p-value estimates for O1 GWTs plus GW170104.}
\end{itemize}
We discuss the comparison of performances with the cWB ES search in section \ref{sec:comparison}.


\begin{table}
\setlength{\tabcolsep}{6pt} 
\renewcommand{\arraystretch}{1.8} 
\begin{tabular}{ l | c | c | c }
    \hline
    \hline
    \multicolumn{4}{ c }{Upper limits on echoes amplitude} \\
    \hline
    GW name & $h_\mathrm{rss}^\mathrm{MR}\cdot\frac{10^{-23}}{\sqrt{Hz}}$ & $h_\mathrm{rss}^\mathrm{UL}\cdot\frac{10^{-23}}{\sqrt{Hz}}$ & $\frac{h_\mathrm{rss}^\mathrm{UL}}{h_\mathrm{rss}^\mathrm{MR}}$ \\
    \hline
GW150914 		& 16.0 & 3.4 &  0.21 \\ 

GW170104 		& 11.5 	& 2.2 &  0.20 \\ 
GW170809 		& 11.3 	& 2.5 &  0.22 \\ 
GW170814 		& 12.1 	& 2.5 &  0.21 \\ 
GW170823 		& 9.8 	& 2.5 &  0.26 \\ 

GW190408\_181802 	& 5.5 	& 1.7 &  0.31 \\ 
GW190412 		& 4.2 	& 1.3 & 0.31 \\ 
GW190513\_205428	& 8.1 	& 1.4 &  0.17 \\ 
GW190521 		& 14.8 	& 2.4 &  0.15 \\ 
GW190521\_074359 	& 15.1 	& 2.5 &  0.17 \\ 
GW190814 		& 2.0 	& 1.5 & 0.75 \\ 
GW190828\_063405 	& 15.0 	& 1.6 & 0.11 \\ 
 
GW191109\_010717 	& 6.2 	& 2.1 & 0.34 \\ 
GW200225\_060421	& 5.1 	& 2.8 & 0.55 \\ 
GW200311\_115853 	& 7.9 	& 2.0 & 0.25 \\ 
    \hline
GW200224\_222234$^{\text{\dag}}$ & 10.8 & 3.7 & 0.34 \\ 
    \hline
    \hline
    \end{tabular}
    \caption{List of the BBH GWTs selected for setting confidence intervals on the echo's amplitude.
    They are a subset of the loudest ones listed in table \ref{tab:event_list}.
    The columns report: the GWT name; the merger-ringdown $h_\mathrm{rss}$ of the primary BBH GWT, $h_\mathrm{rss}^\mathrm{MR}$; the upper limit in terms of $h_\mathrm{rss}$ of possible echo candidates inside the PMW, $h_\mathrm{rss}^\mathrm{UL}$; the relative amplitude upper limit defined as the ratio between the $h_\mathrm{rss}^\mathrm{UL}$ and $h_\mathrm{rss}^\mathrm{MR}$.
    \newline
    \dag: this GWT event is affected by a loud instrumental glitch in the PMW (see appendix \ref{app:gw200224}).}
\label{tab:event_list_upperlimits}
\end{table}

\subsection{Upper limits on \texorpdfstring{$h_\mathrm{rss}$}{} of echoes}
\label{subsec:hrss}
The confidence belt construction procedure requires SIG analyses with extended statistics.
Therefore, we prioritised the GWTs with a merger and ringdown (MR) SNR$_\mathrm{MR} \, \geq 7$, as reported in \cite{tgr3,gwtc3}, if detectable by the cWB all-sky burst search.
We also added to this list the outstanding GW event GW190814 \cite{APJGW190814}.

All confidence intervals result in upper limits on the $h_\mathrm{rss}$ of the echo signals, $h_\mathrm{rss}^\mathrm{UL}$ (see table \ref{tab:event_list_upperlimits}) with the exception of GW200224 (see section \ref{subsec:p-value}, and appendix \ref{app:gw200224}).

Typical upper limits values are in the $h_\mathrm{rss}$ range $\sim 1 \div 4 \times 10^{-23} /\sqrt{Hz}$ at 95$\%$ coverage.
The results in terms of $h_\mathrm{rss}$ can be directly converted to GW strain amplitudes through eq.(\ref{eq:hrss}), once a specific waveform of echo signal is assumed.

The ratios between $h_\mathrm{rss}^\mathrm{UL}$ and the merger-ringdown $h_\mathrm{rss}$ of the primary BBH GWT, $h_\mathrm{rss}^\mathrm{MR}$, are also reported in table \ref{tab:event_list_upperlimits}.
These ratios are our measured amplitude upper limits in relative terms, though their connection to the echo's $A$ parameter (see section \ref{sec:echoes}) depends on the actual morphologies of echo models and of the primary BBH GWT.
In the approximation that the merger-ringdown and each echo pulse share similar morphologies (e.g. similar central frequency and number of cycles), then the reported $h_\mathrm{rss}$ ratios can be considered to be equivalent to upper limits on $A$. 
They are conservative upper limits in case more echo pulses are detected by cWB ES search within the PMW.


\section{Comparison with previous searches for echoes}
\label{sec:comparison}
Here we provide some comments on the performances of the cWB ES search with respect to previously reported methods, being aware, however, that a full comparison of performances would require additional coordinated simulations which are computationally costly and beyond the scope of this paper.
In particular, we are not able to provide comparisons with LVK searches for echoes reported in \cite{tgr1,tgr2,tgr3}, because the published information on this topic is not detailed enough.
Instead, a partial comparison is feasible with a few dedicated papers: we focus on a previous model-independent search using simulated data \cite{tsangPRD98}, and on three template-based searches \cite{ricoPRD99,westerweckPRD97,abedifull}. \newline

\textbf{Model-independent search method by Tsang et al.} \cite{tsangPRD98}.
\label{subsec:tsang}
This general search method for echoes has been first tested on simulated LIGO Hanford and Livingstone detector data assuming Gaussian noise \cite{tsangPRD98}, and then performed a search using real data on GWTs detected in O1 and O2 \cite{tsangPRD101} and in O3 \cite{tgr3}. 
In simulated Gaussian noise, ref. \cite{tsangPRD98} shows that echo signals are confidently detectable above SNR $=12$. 
In addition, at SNR $=8$ the false alarm probability of noise fluctuations misidentified as signals is at the level of a few $\%$. 
The comparison with our cWB ES search can only be semi-quantitative since no information about the detection efficiency as a function of echo parameters is available in \cite{tsangPRD98,tsangPRD101}.
We can point out that for the cWB ES search on real HL data around GW150914, a signal delivering SNR$^{\scaleto{\mathrm{PMW}}{4pt}} = 12$ would also ensure a very confident detection, with a measured $100\%$ detection probability at false alarm probability as low as our measurement limit, $0.1\%$.
Moreover, with SNR$^{\scaleto{\mathrm{PMW}}{4pt}} = 8$ at the selected false alarm probability of $5\%$, the detection probability of cWB ES ranges from $95\%$ to $99\%$ depending on the statistics of noise outliers in different periods of observation. 
This means that cWB ES achieves high detection performances also at SNR$^{\scaleto{\mathrm{PMW}}{4pt}} = 8$ in real noise. 
Moreover, our off-source simulations clearly show that the data are not compliant with a stationary Gaussian noise model in the low SNR range of interest in the proximity of most BBH GWTs. \newline

\textbf{Model-dependent search method by Rico K. L. Lo et al.} \cite{ricoPRD99} \label{subsec:rico}
This has been the first model-dependent search that challenged the claim of an echo discovery after GW150914 by Abedi \cite{abediPRD96}. 
Figure 4 from \cite{ricoPRD99} shows that an A parameter greater than 0.3 can be detected with a 5$\sigma$ threshold for the GW150914 emission in Gaussian noise and using Advanced LIGO design sensitivity.
The signal model’s parameters used in \cite{ricoPRD99} for this result are very similar to the ones used here: the only non-negligible differences are on $\gamma$ parameter and number of pulses.
In \cite{ricoPRD99}, three echo-like pulses have been injected with $\gamma \sim$ 0.9, while here we injected two with $\gamma = 0.5$.
Detectability of A $\sim 0.3$ is well in the ballpark of our method when using real data, see last column from table \ref{tab:event_list_upperlimits}.
In particular, for GW150914, our search constrains A below 0.21 with 95$\%$ confidence.
Moreover, figure \ref{fig:belt} shows that cWB ES search can identify echo signals at 95$\%$ confidence when SNR$^{\scaleto{\mathrm{PMW}}{4pt}}_{rec}$ passes the 4 $\sigma$ threshold, a performance which is comparable with what reported in table 3 from \cite{ricoPRD99}
\footnote{In table 3 of \cite{ricoPRD99}, the threshold on the detection statistics which corresponds to 4 $\sigma$ in Gaussian noise, is delivering a 2 $\sigma$ confidence in real GW150914 noise, similarly to our result.}. \newline

\textbf{Model-dependent search method by Westerweck et al.} \cite{westerweckPRD97}
\label{subsec:julian}
This template-based search has been deployed on real data analyzing four BBH GWTs (including GW150914) and does not find violations of the null hypothesis.
It estimates the p-values of results by using different noise instantiations close to the GWTs times, which is a similar method to our BKG analysis.
Instead, the sensitivity of this search is assessed by injecting echo waveforms on simulated Gaussian noise which preserves the actual power spectral density of the LIGO detectors at the GWTs detections.
Figure 2 and 5 in \cite{westerweckPRD97} show that peak amplitudes of echoes detach from the 
noise fluctuations starting from $h_\mathrm{p} \simeq 2 \cdot 10^{-22}$.
In actual noise, our search achieves $50\%$ detection probability with a false alarm probability of $5\%$ for a peak amplitude of the assumed echo waveform $h_\mathrm{p} \sim 2.3  \cdot  10^{-22}$ for GW150914, as estimated from our more general result in terms of $h_{\mathrm{rss}50\%}$ (see Tab.\ref{tab:event_list}).
Therefore, we conclude that the sensitivity of the cWB ES search is at least competitive to that of this template-based search on this specific echo model.
We remark that the implementation of the model-dependent search uses a template bank and requires subtraction of the detected GWT trigger from data prior to matched filtering for the template bank. Such steps add complexity with respect to the cWB ES search. \newline

\textbf{Model-dependent analysis by Abedi.} \cite{abedifull}
\label{subsec:abedi65}
Another systematic search for a specific echo model has been very recently reported by Abedi \cite{abedifull}.
This search analyses 65 GWTs from the LVK catalog of compact binary coalescences.
The method assumes Gaussian noise close to each GW event.
The main result reported is an upper limit value to the echo amplitude, A, resulting to be A $\leq 0.42$ with a 90\% credible interval, under the assumption that A is equal for all analyzed events.
In addition, the Bayes factor reported for GW190521 stands out as an outlier, suggesting a preference for post-merger echoes rather than the null hypothesis.
In our study, GW190521 shows an on-source p-value equal to $0.569 \pm 0.006$, suggesting that the data in the PMW are compatible with noise.
Moreover, our relative upper limit on the amplitude $h_\mathrm{rss}$ ratio at 95\% coverage is $0.23$ for GW190521 and is as low as $0.13$ for the loudest GWTs.


\section{Conclusions}
\label{sec:conclusions}
This paper describes a search for secondary gravitational wave transients of generic morphology which may occur shortly after the ringdown phase of a primary signal from a Compact Binary Coalescence.
The analysis method is developed on top of the coherent WaveBurst pipeline: it uses the primary GWT as a trigger and follows up the coherent response of the interferometric gravitational wave detectors on a selectable time window, defined with respect to the merger time.

The scientific motivation for this work is the search for gravitational wave echoes after binary black hole mergers.
Such echoes are expected if the final remnant object is not a standard black hole from the general relativity theory, either because the event horizon is not fully absorbing or because the remnant is an exotic compact object larger than the would-be event horizon.
The detection performances of the current search are described in terms of $h_\mathrm{rss}$ strain amplitude and are rather independent of the signal waveform and spectra within a wide signal class.
Therefore, as long as any echo pulse occurs inside the selected time window, from 0.05 to 0.35 $\si{\second}$ or from 0.2 to 1.2 $\si{\second}$ after the merger, the reported results can be interpreted in terms of any echoes' model.

The analysis of the loudest 33 BBH mergers detected during the O1, O2, and O3 observing runs by the LIGO, Virgo, and KAGRA collaborations is consistent with null results (see table \ref{tab:event_list}), so no evidence of echo signals is found.
This search provides separate results for single BBH mergers.
The off-source characterization of the detection efficiency vs false alarm probability and the estimation of p-values of candidates is performed using thousands of real detector noise instantiations.
Therefore, the results do not rely on an a priori noise model and point out that the actual noise statistic is far from Gaussian in most cases, even at low SNR.
The search also provides a morphological reconstruction of candidates and, for the first time, the confidence intervals on the $h_\mathrm{rss}$ amplitude of gravitational wave echoes.
The latters turn out to be upper limits, typically ranging in the interval $\sim 1 \div 4 \times 10^{-23} /\sqrt{Hz}$ in terms of $h_\mathrm{rss}$ (see table \ref{tab:event_list_upperlimits}).

The two loudest candidates found occur after GW190701 and GW200224.
These candidates are also the only ones featuring low enough p-values to require further follow-up investigations.
Their morphological reconstruction clearly points to the dominating presence of known pulsating instrumental noise disturbances at low frequencies, occurring in both the LIGO detectors, and they are by far inconsistent with any published model of echoes.
The pseudo-regular cadence of these disturbances is the likely cause of a systematic error in our initial p-values estimates.

To our knowledge, this search for echoes is delivering the highest sensitivity to the possible presence of gravitational wave echoes occurring within a selected post-merger time window, without relying on signal templates.
Our a posteriori use of morphological information to reject or accept candidates is still a sub-optimal strategy.
An a priori exploitation of loose morphological priors of echo signals will likely improve the current method.

We plan to extend this method also to investigate the post-merger emission after BNS (see e.g. \cite{bnsPRL119}) and NSBH coalescence over a wider frequency range, exploiting the entire spectral sensitivity of the LVK detectors.
Remarkably, this search can be adapted to study other science cases of interest in current GW astronomy, which share the expectation of a  weak GW feature close to the coalescence time of the primary CBC GWT signal.
Examples include investigation of memory effects \cite{chrisPRL67,favataCQG27_memory,hubnerPRD101,hubnerPRD104,tiwariPRD104}, precursors to highly eccentric BBHs \cite{yoshinataPRD103,shubhanshuPRD101,ebbh_LIGO,rossellaNATURE7,ebersolPRD106}, or micro-lensing effects \cite{mariaPRD103_lensing,LVlensingAPJ923}.


\begin{acknowledgments}
The authors would like to thank Andrea Maselli, Francesco Salemi, and Patrick Sutton for their constructive inputs.
We also acknowledge useful discussions with Sophie Bini, Alessandro Martini, and Andrea Virtuoso.
This research used data, software, and web tools from the Gravitational Wave Open Science Center, a service of LIGO Laboratory, the LIGO Scientific Collaboration, and the Virgo Collaboration.
This material is based upon work supported by NSF's LIGO Laboratory which is a major facility fully funded by the National Science Foundation.
Virgo is funded by the French Centre National de Recherche Scientifique (CNRS), the Italian Istituto Nazionale di Fisica Nucleare (INFN), and the Dutch Nikhef, with contributions by Polish and Hungarian institutes.
The authors are grateful for computational resources provided by the LIGO Laboratory and supported by National Science Foundation Grants PHY-0757058 and PHY-0823459.
Andrea Miani thankfully acknowledges the grant provided by the EGO Consortium EGO-DIR-56-2021 and the University of Trento.
Shubhanshu Tiwari is supported by the Swiss National Science Foundation (SNSF) Ambizione Grant Number: PZ00P2-202204.
\end{acknowledgments}


\newpage
\appendix
\section{cWB all-sky burst search vs ES search}
\label{app:th_summary}
Table \ref{tab:parameter} lists cWB \cite{cwbSoftX,CWB} parameters (first column) and their threshold values that can be tuned in a cWB search, comparing the configuration for the cWB all-sky burst search \cite{allsky_o1,allsky_o2,allsky_o3} (second column) to that of the cWB ES search (third column).
The different tuning of $\eta_\mathrm{c}$, T$_\mathrm{gap}$, and SUBRHO thresholds is motivated by the need to grasp lower SNR triggers in the search for echoes, while keeping under control the false alarms.
Additional configuration parameters are defined in the cWB ES search: the time width of the PMW, $\Delta t^{\scaleto{\mathrm{PMW}}{4pt}}$; a blind time after the coalescence time, $t_\mathrm{blind}$; the fraction of correlated energy in the PMW, $c^{\scaleto{\mathrm{PMW}}{4pt}}_\mathrm{c}$.

\begin{table}
	\centering
	\setlength{\tabcolsep}{6pt} 
	\renewcommand{\arraystretch}{1.5} 
    \begin{tabular}{ l | c | c }
  	\hline
  	\hline
  	\multicolumn{3}{ c }{configuration parameters} \\
  	\hline
  	Parameters & All-Sky O3 search & ES search \\
    \hline
    bpp & 0.001 & 0.001 \\ 
    subnet & 0.5 & 0.5 \\ 
    $c_\mathrm{c}$ & 0.5 & 0.5 \\ 
    $\eta_\mathrm{c}$ & 5.0 & 3.5 \\ 
    $A_\mathrm{core}$ & 1.7 & 1.7 \\ 
    $T_\mathrm{gap}$ & 0.2 $\si{\second}$ & 2.0 $\si{\second}$ \\ 
    $F_\mathrm{gap}$ & 128.0 $\si{\hertz}$ & 128.0 $\si{\hertz}$ \\ 
    SUBRHO & 5.5 & 3.5 \\ 
    SUBNET & 0.1 & 0.1 \\ 
    PMW & not used & $\Delta t^{\scaleto{\mathrm{PMW}}{4pt}} = \SI{1}{\second}$ \\ 
    $t_\mathrm{blind}$ & not used & $t_\mathrm{blind} \geq \SI{40}{\milli\second}$ \\ 
    $c^{\scaleto{\mathrm{PMW}}{4pt}}_\mathrm{c}$ & not used & $c^{\scaleto{\mathrm{PMW}}{4pt}}_\mathrm{c} \geq 0.5$ \\
    \hline
    \hline
    \end{tabular}
    \caption{Comparison of the cWB main production thresholds between all-sky burst search \cite{allsky_o3} (second column) and ES search (third column).}
    \label{tab:parameter}
\end{table}

\begin{figure}
    \centering
    \begin{subfigure}{0.45\textwidth}
		\centering
        \includegraphics[width=\textwidth]{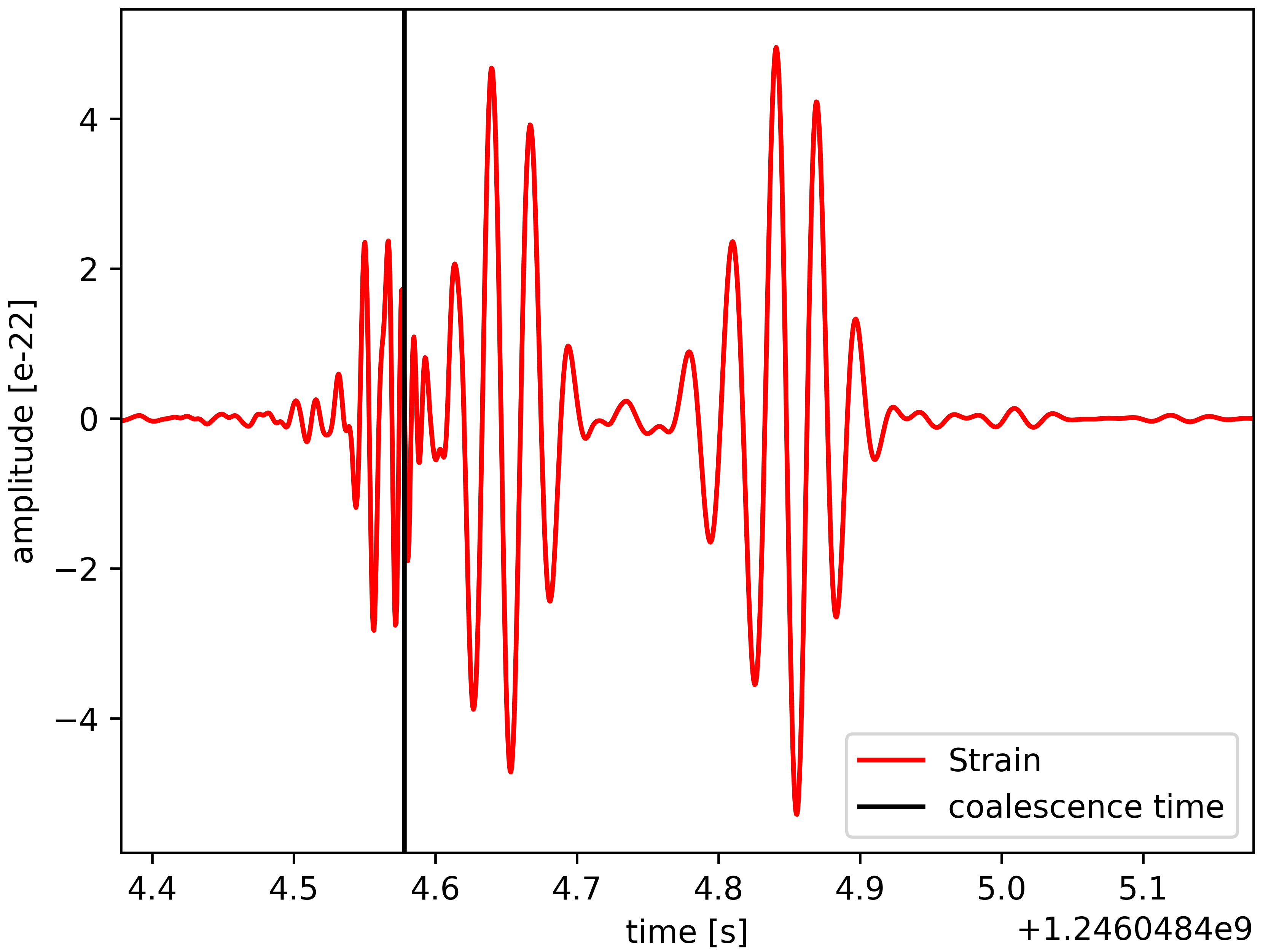}
        \caption{GW190701 - L1 strain.}
        \label{fig:gw190701_wf_str}
	\end{subfigure}
    \hfill
    \begin{subfigure}{0.45\textwidth}
		\centering
		\includegraphics[width=\textwidth]{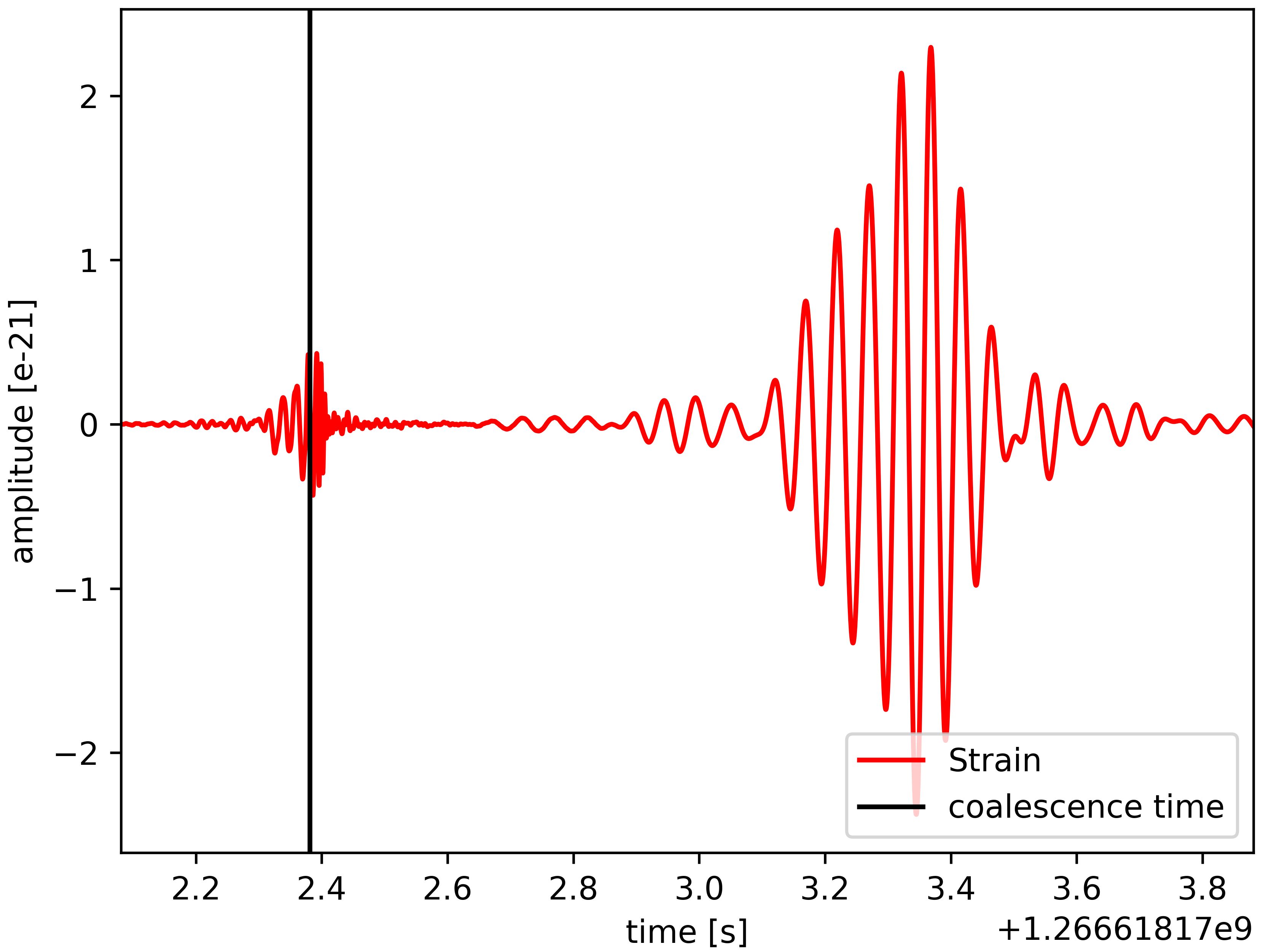}
		\caption{GW200224 - L1 strain.}
        \label{fig:gw200224_wf_str}
    \end{subfigure}
    \caption{On top: Fig. \ref{fig:gw190701_wf_str} shows the strain amplitude waveform of GW190701 and its post-merger as function of time for L1 detector.
    On the bottom: Fig. \ref{fig:gw200224_wf_str} reports the strain amplitude of GW200224 and its post-merger as function of time for L1 detector.
    In both plots the black vertical line marks the coalescence time of the binary.
    For both the GW signals are clearly visible the glitches providing the energy excesses when running the ES search.}
\label{fig:gw_strain}
\end{figure}

\section{Followup of loudest candidates}
\label{app:special_ev}
From the analysis of the p-values of the BBH GWTs (see section \ref{subsec:p-value}), two events are selected for deeper investigations since they are consistent with a FDR $\leq 10\%$: GW190701 and GW200224.
Estimating the p-values on a different, more local set of noise instantiations results in higher p-values, which points to some systematic bias in our estimating procedure.
Nevertheless, these two local p-values are still the only ones $\leq 1\%$, further motivating the following deeper investigations on GW190701 and GW200224.

The morphological study of the PMW on-source event allows to gather information about the reconstructed SNR of the energy excess, its arrival time, mean frequency, and the reconstructed waveform.
Additional tests have been deployed as well, like performing a single detector analysis of the on-source morphology, with the subtraction of the primary BBH waveform.
The information of the morphological studies are then compared with the theoretical expectation of echo models (see section \ref{sec:echoes}) and with the known noise disturbances.

\begin{figure*}
    \begin{subfigure}{0.30\textwidth}
		\centering
        \includegraphics[width=\textwidth]{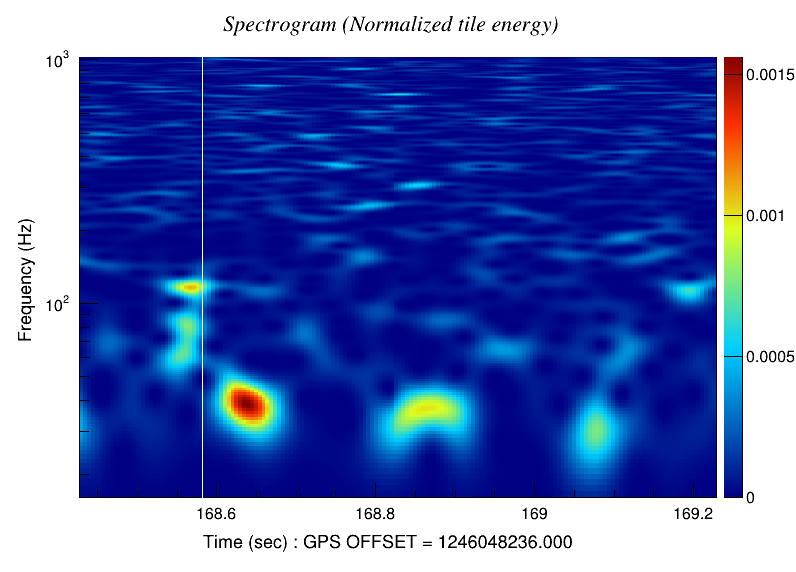}
        \caption{L1 TF map.}
        \label{fig:gw190701_TF_L}
	\end{subfigure}
    \hfill
    \begin{subfigure}{0.30\textwidth}
		\centering
		\includegraphics[width=\textwidth]{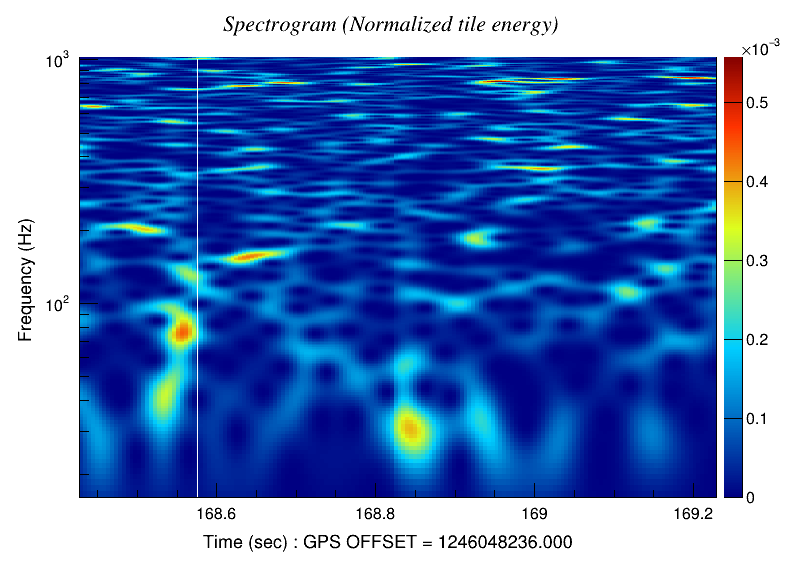}
		\caption{H1 TF map.}
        \label{fig:gw190701_TF_H}
    \end{subfigure}
    \hfill
    \begin{subfigure}{0.30\textwidth}
		\centering
		\includegraphics[width=\textwidth]{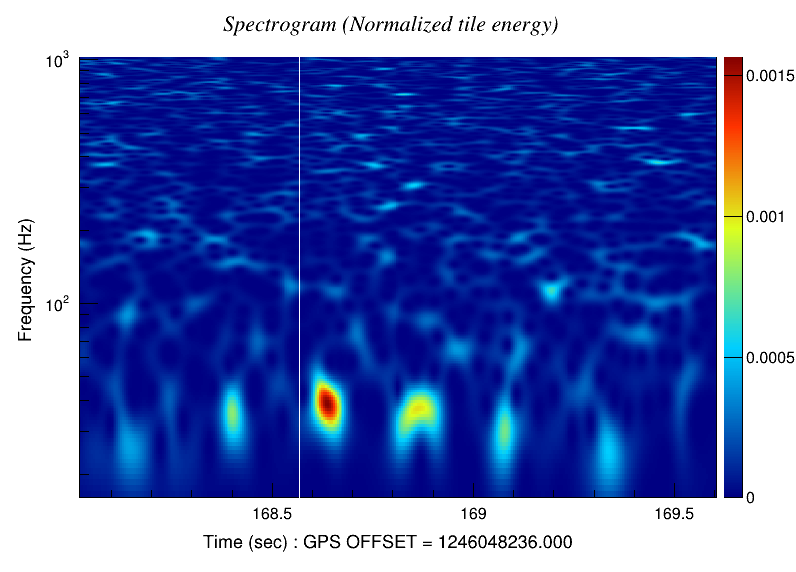}
		\caption{L1 TF map - BBH sub.}
        \label{fig:gw190701_TF_L_sub}
    \end{subfigure}
    \vskip
    \baselineskip
    \begin{subfigure}{0.30\textwidth}
    	\centering
        \includegraphics[width=\textwidth]{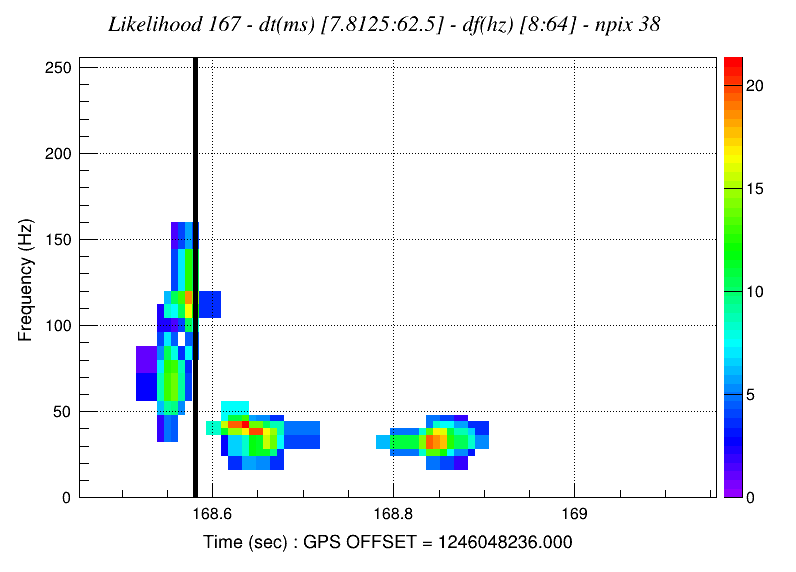}
        \caption{LH - Likelihood.}
        \label{fig:gw190701_lik_LH}
    \end{subfigure}
    \hfill
    \begin{subfigure}{0.30\textwidth}
    	\centering
        \label{fig:gw190701_lik_LH_sub}
    \end{subfigure}
    \begin{subfigure}{0.30\textwidth}
		\centering
        \includegraphics[width=\textwidth]{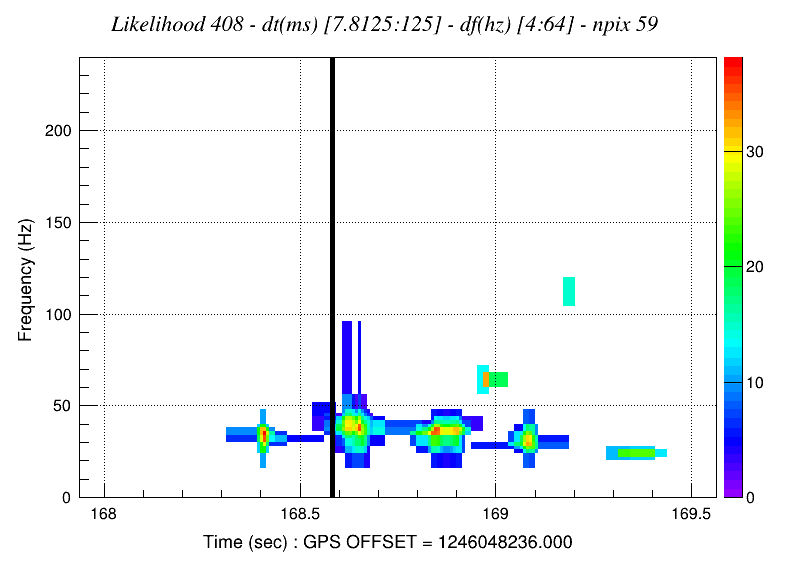}
        \caption{L - Likelihood - BBH sub.}
        \label{fig:gw190701_lik_L_sub}
	\end{subfigure}
    \hfill
    \caption{In this figure: plots \ref{fig:gw190701_TF_L} and \ref{fig:gw190701_TF_H} show the time-frequency map of GW190701 event in L1 and H1 detector respectively.
    Plot \ref{fig:gw190701_TF_L_sub} shows the event in L1 detector once the best template of GW190701 between its posterior samples is subtracted from the data.
    The white vertical line marks the coalescence time of the BBH event.
    Plot \ref{fig:gw190701_lik_LH} show the reconstructed maximum likelihood of the event for the LH network, while plot \ref{fig:gw190701_lik_L_sub} display the same quantity but for a single detector search (L1) and after the best GW190701 template is subtracted from the data.
    The black vertical line marks the coalescence time of the BBH event.
    Note that the color bars scale between L and H detectors have a different range.}
\label{fig:gw190701}
\end{figure*}

\subsection{GW190701}
\label{app:gw190701}
Figure \ref{fig:gw190701_wf_str}, shows the reconstructed strain signal waveform of GW190701 in L1 detector.
Here, the BBH signal is the smallest bump on the left while the two bumps on its right are the post-merger energy excesses.
Among them, the most interesting one is the second (at time $\sim \SI{168.86}{\second}$) since it is the one falling inside our PMW.
The post-merger candidate shows higher strain, and longer time duration, around $\geq \SI{100}{\milli\second}$, with respect to the BBH event, and no echo models are consistent, to our knowledge, with these features.

The entire on-source event (BBH + PM signals) has an overall SNR content around SNR $\sim 12.9$ with a $\rho \sim 4.8$, and $c_\mathrm{c} \sim 0.57$ that is an unusually low value for an event with such an SNR.
Figures \ref{fig:gw190701_TF_L} and \ref{fig:gw190701_TF_H}, the two TF maps of the event for each detector, show that in L1-detector, after the BBH event, there are three post-merger energy excesses, at times $\sim \SI{168.64}{\second}$, $\sim \SI{168.86}{\second}$, and $\SI{169.07}{\second}$, while in the post-merger of H1 there is only one clear energy excess at $\sim \SI{168.84}{\second}$.
This energy distribution asymmetry explains the low value of the correlation coefficient $c_\mathrm{c}$ suggesting that a noise realisation is a preferred explanation for such an observation since it does not match up with echo signal predictions.
Furthermore, in the bottom row of figure \ref{fig:gw190701} there is the network on-source likelihood \ref{fig:gw190701_lik_LH} TF map.
At  time $\sim\SI{168.55}{\second}$ there is the chirping cluster of pixels representing the GW190701 event, going from frequencies around $\sim\SI{40}{\hertz}$ up to $\sim\SI{150}{\hertz}$, while in the post-merger, at the time $\sim\SI{168.84}{\second}$, is clearly visible the energy excess.
It has a central frequency around $f_0 \sim [30-40] \, \si{\hertz}$ which is not a frequency range expected for echoes: they should possess similar frequencies or higher than the BBH merger one \cite{wangPRD97}.
Finally, figure \ref{fig:gw190701_lik_L_sub} shows the on-source likelihood TF map after the BBH subtraction for the single L1 detector configuration.
A repetition of similar pulses is visible both before and after GW190701 coalescence time ($\sim\SI{168.55}{\second}$).
This is again inconsistent with echo models, and points to an accidental coincidence with noisy features polluting L1 data.

\begin{figure*}
    \begin{subfigure}{0.30\textwidth}
		\centering
        \includegraphics[width=\textwidth]{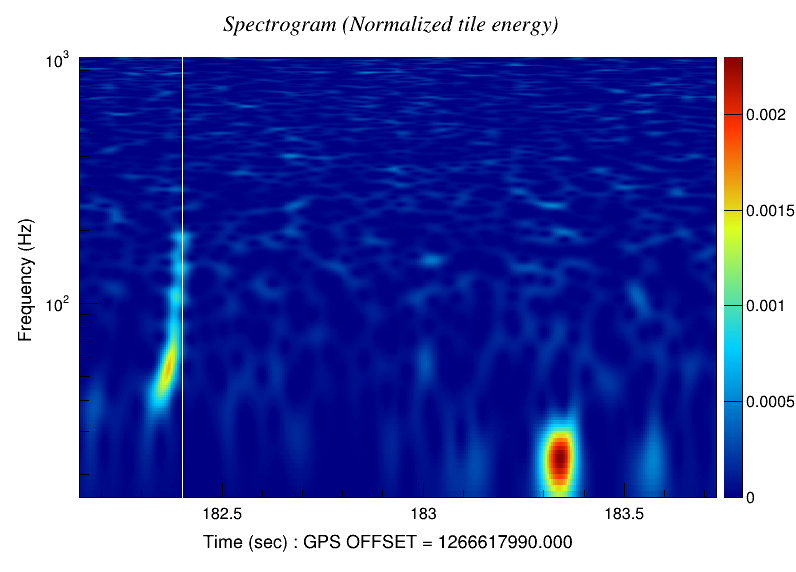}
        \caption{L1 TF map.}
        \label{fig:gw200224_TF_L}
	\end{subfigure}
    \hfill
    \begin{subfigure}{0.30\textwidth}
		\centering
		\includegraphics[width=\textwidth]{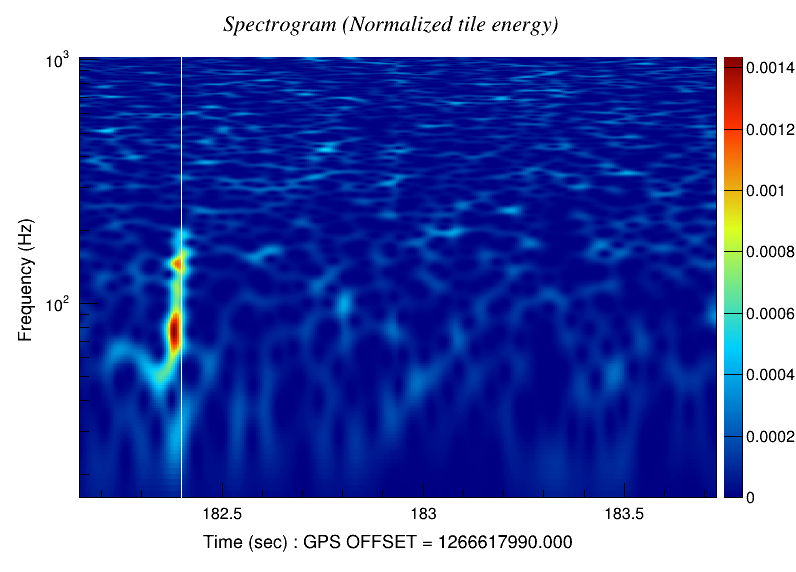}
		\caption{H1 TF map.}
        \label{fig:gw200224_TF_H}
    \end{subfigure}
    \hfill
    \begin{subfigure}{0.30\textwidth}
		\centering
		\includegraphics[width=\textwidth]{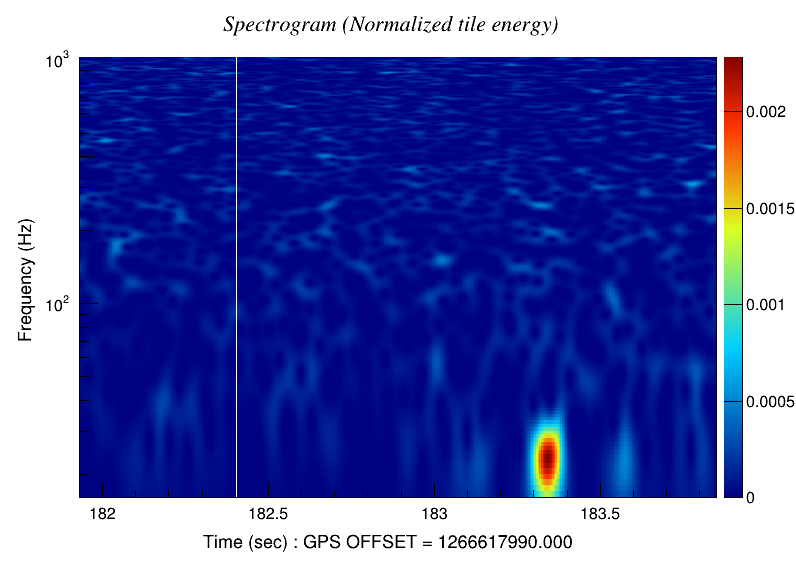}
		\caption{L1 TF map - BBH sub.}
        \label{fig:gw200224_TF_L_sub}
    \end{subfigure}
    \vskip
    \baselineskip
    \begin{subfigure}{0.30\textwidth}
    	\centering
        \includegraphics[width=\textwidth]{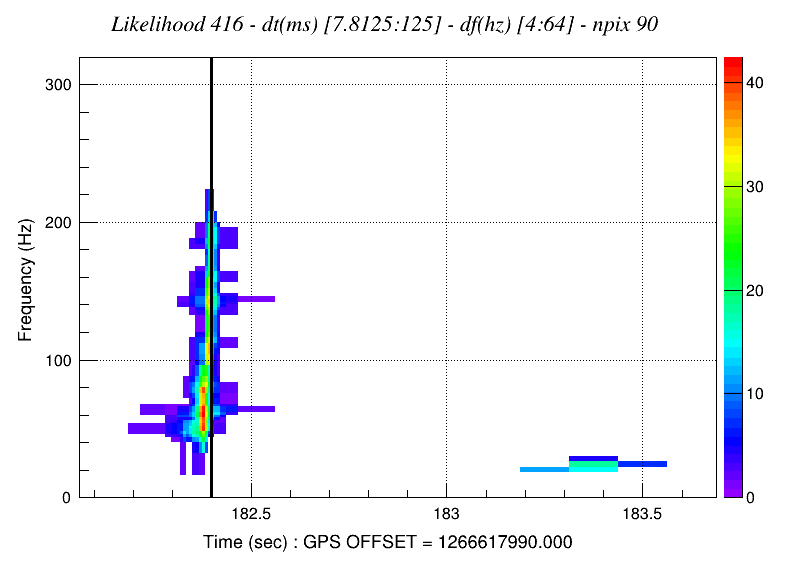}
        \caption{LH - Likelihood.}
        \label{fig:gw200224_lik_LH}
    \end{subfigure}
    \hfill
    \begin{subfigure}{0.30\textwidth}
    	\centering
        \label{fig:gw200224_lik_LH_sub}
    \end{subfigure}
    \begin{subfigure}{0.30\textwidth}
		\centering
        \includegraphics[width=\textwidth]{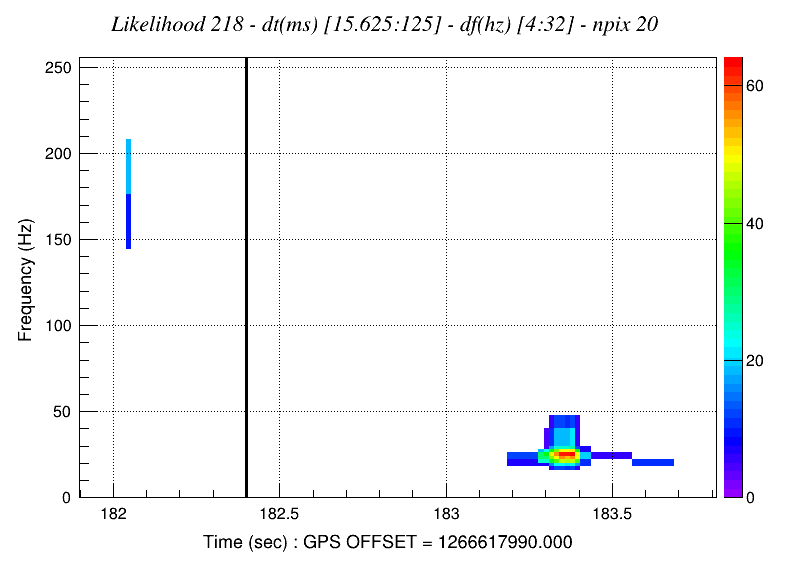}
        \caption{L - Likelihood - BBH sub.}
        \label{fig:gw200224_lik_L_sub}
	\end{subfigure}
    \hfill
    \caption{In this figure: plots \ref{fig:gw200224_TF_L} and \ref{fig:gw200224_TF_H} show the time-frequency map of GW200224 event in L1 and H1 detector respectively.
    Plot \ref{fig:gw200224_TF_L_sub} shows the event in L1 detector once the best template of GW200224 between its posterior samples is subtracted from the data.
    The white vertical line marks the coalescence time of the BBH event.
    Plot \ref{fig:gw200224_lik_LH} show the reconstructed maximum likelihood of the event for the LH network, while plot \ref{fig:gw200224_lik_L_sub} display the same quantity but for a single detector search (L1) and after the best GW200224 template is subtracted from the data.
    The black vertical line marks the coalescence time of the BBH event.
    Note that the color bars scale between L and H detectors have a different range.}
\label{fig:gw200224}
\end{figure*}

\subsection{GW200224}
\label{app:gw200224}
To study the post-merger on-source energy excess detected in GW200224 we deploy the same strategy in GW190701.
Figure \ref{fig:gw200224_wf_str} shows the on-source strain waveform of the entire event, with the BBH being the small signal on the left.
The time duration of the PM signal, $\sim \SI{400}{\milli\second}$ as well as its time distance of $\sim \SI{1}{\second}$ to the merger time of the BBH do not match the theoretical predictions of echo signals.
Following eq.(\ref{eq:dt_echo_M30}) $t_\mathrm{echo}$ is predicted to be $\sim \SI{}{\milli\second}$ after $t_\mathrm{coa}$.
Moreover, the TF map of the on-source event, figure \ref{fig:gw200224_TF_L} shows that the mean frequency of this PM excess of energy is around $\SI{40}{\hertz}$ well below the expected frequency values for echoes.

Figures \ref{fig:gw200224_TF_L} and \ref{fig:gw200224_TF_H}, the TF maps of the event in L1 and H1 respectively, shows that the PM signal is present only in L1 detector, while in H1 such high energy excess is not reconstructed.
Since the two LIGO detectors are nearly aligned and are sensitive to the same GW's polarisation, for real astrophysical events such energy imbalance in the detectors is suspicious.

We proceed in subtracting to GW200224 on-source event the best PE model describing that same BBH event, then on the subtracted data we run the single detector ES search.
The result is displayed in figure \ref{fig:gw200224_lik_L_sub}.
Here, no undetected energy excess other than the investigated one appears, suggesting that we are not in a scenario similar to the single detector analysis of GW190701.
The energy outlier has a SNR $\sim 10.4$, while the overall SNR of the BBH signal plus post-merger excess of energy is equal to $\sim 16.8$ (in single detector mode).

\begin{figure*}
    \centering
    \includegraphics[width=0.48\textwidth]{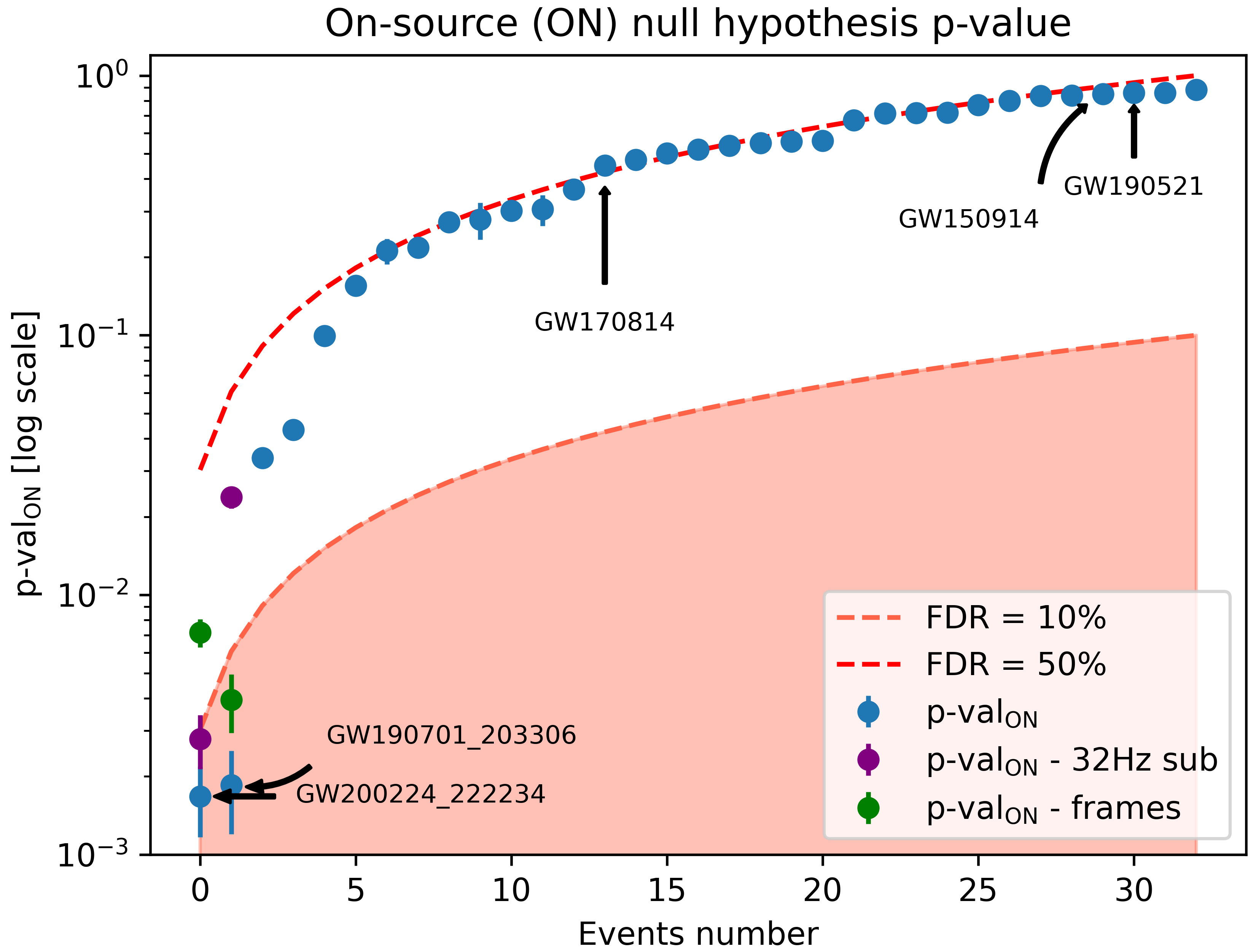}
    \caption{Same plot of figure \ref{fig:pvalue}, where the \textbf{green dots} are the on-source p-value obtained from the ES search carried on using only the $\SI{4096}{\second}$ around the main BBH event.
The \textbf{violet dots} are the on-source p-value for the ES search performed over the same standard analysis data but applying a $\SI{32}{\hertz}$ data mitigation plugin to suppress some noisy contaminating the O3a and O3b observational periods.}
    \label{fig:pvalue_update}
\end{figure*}

\subsection{cWB ES search with 32 Hz mitigation}
\label{app:32Hz}
The PMW on-source morphologies hint to a possible data pollution by a glitch family identified in the frequency range $\in \left(16,40\right) \, \si{\hertz}$ \cite{caberoCQG32Hz,DQo1}.
Therefore, we repeated the ES search for these two GWTs by including a specific single detector data filter \cite{32HzPRD103,regression32Hz} that estimates the power oscillations within the frequency range $\in\left(16-40\right) \, \si{\hertz}$ and attenuates them.
We label such analysis as $\SI{32}{\hertz}$-ES search, to differentiate it from the standard ES search.
The measured on-source null hypothesis p-values when the noise around $\SI{32}{\hertz}$ frequencies is mitigated result:
\begingroup
\addtolength{\jot}{1.5em}
\begin{align}
	&\text{GW190701:} \quad  \text{p-value}_{\text{ON}}^{\SI{32}{\hertz}} = 0.024 \pm 0.002
	\label{eq:gw190701_pval_32} \\
	&\text{GW200224:} \quad  \text{p-value}_{\text{ON}}^{\SI{32}{\hertz}} = 0.003 \pm 0.001
	\label{eq:gw200224_pval_32} \,
\end{align}
\endgroup
and they are plotted in figure \ref{fig:pvalue_update} as the violet dots.
This noise mitigation rules out the post-merger event candidate GW190701, while for the PM of GW200224 the p-value is still within the FRD$\leq 10\%$.

This study together with the morphological investigation of the PMW energy excesses of GW190701 and GW200224 (see appendix \ref{app:gw190701} and \ref{app:gw200224}), show that it is reasonable to assume them to be non stationary noise feature polluting the data and especially affecting L detector.
These noise transients posses a central frequency around $(30,40) \si{\hertz}$, and have a greater time duration ($\sim$ hundred of $\si{\milli\second}$) with respect to the expected one for echo signals ($\sim$ tens of $\si{\milli\second}$), so around one order of magnitude bigger.


\newpage
\bibliography{bibliography}

\end{document}